\documentclass[
  aps,
  prb,
  reprint,
  superscriptaddress,
  amsmath,amssymb,
]{revtex4-2}
\usepackage{graphicx}
\usepackage{bm}
\usepackage{siunitx}
\usepackage{xcolor}
\usepackage{hyperref}
\usepackage{natbib}
\usepackage{braket}
\usepackage{layouts}

\usepackage{mymacros}

\newcommand{\figpath}{figures}

\begin{document}

\title{Flux-tunable Josephson Effect in a Four-Terminal Junction}

\author{Christian G. Prosko}
\affiliation{QuTech and Kavli Institute of Nanoscience, Delft University of Technology, 2600 GA Delft, The Netherlands}

\author{Wietze D. Huisman}
\affiliation{QuTech and Kavli Institute of Nanoscience, Delft University of Technology, 2600 GA Delft, The Netherlands}

\author{Ivan Kulesh}
\affiliation{QuTech and Kavli Institute of Nanoscience, Delft University of Technology, 2600 GA Delft, The Netherlands}

\author{Di Xiao}
\affiliation{Department of Physics and Astronomy, Purdue University, West Lafayette, Indiana 47907, USA}

\author{Candice Thomas}
\affiliation{Department of Physics and Astronomy, Purdue University, West Lafayette, Indiana 47907, USA}

\author{Michael J. Manfra}
\affiliation{Department of Physics and Astronomy, Purdue University, West Lafayette, Indiana 47907, USA}
\affiliation{School of Materials Engineering, Purdue University, West Lafayette, Indiana 47907, USA}
\affiliation{Elmore School of Electrical and Computer Engineering, Purdue University, West Lafayette, Indiana 47907, USA}

\author{Srijit Goswami}
\email{S.Goswami@tudelft.nl}
\affiliation{QuTech and Kavli Institute of Nanoscience, Delft University of Technology, 2600 GA Delft, The Netherlands}

\date{\today}

\begin{abstract}
  We study a phase-tunable four-terminal Josephson junction formed in an InSbAs two-dimensional electron gas proximitized by aluminum.
  By embedding the two pairs of junction terminals in asymmetric DC SQUIDs we can control the superconducting phase difference across each pair, thereby gaining information about their current-phase relation.
  Using a current-bias line to locally control the magnetic flux through one SQUID, we measure a nonlocal Josephson effect, whereby the current-phase relation across two terminals in the junction is strongly dependent on the superconducting phase difference across two completely different terminals.
  In particular, each pair behaves as a  $\phi_0$-junction with a phase offset tuned by the phase difference across the other junction terminals.
  Lastly, we demonstrate that the behavior of an array of two-terminal junctions replicates most features of the current-phase relation of different multiterminal junctions.
  This highlights that these signatures alone are not sufficient evidence of true multiterminal Josephson effects arising from hybridization of Andreev bound states in the junction.
\end{abstract}

\maketitle

\section{Introduction}
\label{sec:4termjj_introduction}

Multiterminal Josephson junctions (JJs) formed from more than two terminals
have current-phase relations determined by the superconducting phases of all terminals \cite{Beenakker_1991}.
The Andreev bound state (ABS) spectrum of multiterminal junctions can manifest topological phases containing Majorana bound states \cite{Van_Heck_2014} or protected Weyl nodes in their band structure \cite{Riwar_2016,Xie_2017,Houzet_2019,Fatemi_2021,Barakov_2023}, with the superconducting phases of the terminals behaving as momentum degrees of freedom \cite{Riwar_2016}.
To form Weyl nodes in the absence of a flux through the junction \cite{Meyer_2017} at least four terminals are required, because an $n$-terminal junction manifests topology in $n-1$ dimensions.
Additionally, four-terminal JJs (4TJJs) are expected to exhibit non-trivial current-phase relations (CPRs) of the supercurrent through the junction \cite{de_Bruyn_Ouboter_1995,Zareyan_1999,Omelyanchouk_2000,Amin_2001,Alidoust_2012}, and can form a superconducting phase qubit bypassing some constraints of conventional flux qubits \cite{Amin_2002}.
In particular, the phase difference across two terminals is expected to induce a phase difference and supercurrent across the other two terminals, giving them applicability as switching elements for superconducting electronics \cite{Alidoust_2012}.

Previous work on multiterminal JJs observed multiterminal DC and AC Josephson effects \cite{Vleeming_1996,Vleeming_1999,Draelos_2019,Pankratova_2020,Graziano_2020,Arnault_2021,Graziano_2022,Lee_2022_thesis,Chiles_2023}, signatures of supercurrent mediated by Cooper quartets \cite{Pfeffer_2014,Cohen_2018,Huang_2022,Zhang_2023}, and strong diode behavior \cite{Gupta_2023,Chiles_2023}.
For three-terminal JJs, signatures of Andreev molecules \cite{Coraiola_2023,Matsuo_2023_ArXiv_molecule} and more complicated subgap state spectra affected by spin-orbit coupling have been observed \cite{Coraiola_2023_ArXiv_spin}.
Meanwhile, despite numerous experiments on 4TJJs \cite{Vleeming_1999,Draelos_2019,Pankratova_2020,Huang_2022,Zhang_2023},
the CPR of any four-terminal junction has yet to be probed with control over two or more phase degrees of freedom.

We thus consider a four-terminal JJ (4TJJ) embedded in two asymmetric DC SQUIDs penetrated by independently controllable magnetic fluxes.
This allows us to control two phase differences across pairs of terminals in the junction.
Accordingly, we can measure SQUID oscillations containing information about the CPR across their corresponding 4TJJ terminals in the form of a current-flux relation (CFR).
Furthermore, with four instead of three terminals we are able to measure a Josephson effect which is fully `nonlocal' in that the CPR between two superconducting terminals is modified by a phase difference across a completely independent pair of terminals \cite{Omelyanchouk_2000,Amin_2001}.
Correspondingly, two terminals of the 4TJJ form a tunable $\phi_0$-junction with a phase offset tunable in a range larger than $0.2\Phi_0$ where $\Phi_0=h/2e$ is the superconducting flux quantum \cite{Buzdin_2008}.
For this experiment and others in three-terminal JJs \cite{Haxell_2023,Matsuo_2023_ArXiv_nonlocal}, we model the junction as an array of two-terminal junctions and find that this $\phi_0$-junction effect can exist even in the absence of a hybridized ABS spectrum in the junction, necessitating other experimental signatures for ABS hybridization.

\section{Device Design \& Characterization}
\label{sec:4termjj_design}

\begin{figure*}[t!]
  \centering
  \includegraphics{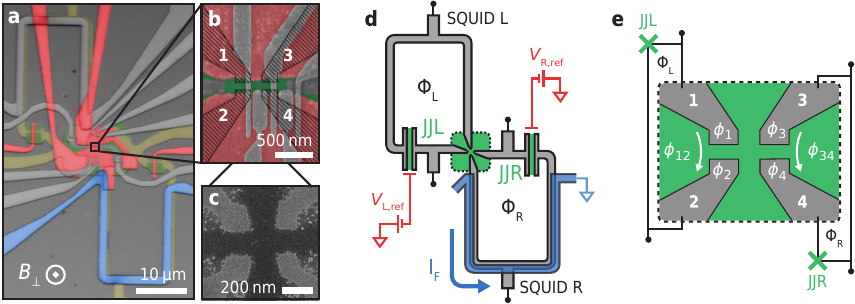}
  \caption{
    Experimental Setup.
    \textbf{(a)} False-color optical image of measured Device A, highlighting the 4TJJ's depletion gate and reference junction gates (red), the flux-bias line (blue), exposed portions of the 2DEG mesa (green), and the Al forming the SQUID circuits (yellow).
    \textbf{(b)} False-color scanning electron micrograph of Device A near its central 4TJJ.
    Unused misaligned gates in the second layer and unused normal-metal tunnel probe leads are shown in gray, while the designed pattern of the Al 4TJJ is superimposed in black on the image.
    \textbf{(c)} As in \textbf{(b)} but before any gates were deposited.
    Overetching of the Al caused the 4TJJ junction dimensions to be slightly larger than the design shown in \textbf{(b)}.
    Residues are visible surrounding the designed pattern, arising during an Al etching step.
    \textbf{(d)} Schematic of the device consisting of SQUIDs labeled L and R penetrated by fluxes $\Phi_\mathrm{L/R}$.
    \textbf{(e)} Schematic of the 4TJJ itself, with the superconducting terminals numbered, and their wave function phases and phase differences labeled.
  }
  \label{fig:4termjj_device}
\end{figure*}

The devices (A and B) are formed in an InSb$_{0.86}$As$_{0.14}$ two-dimensional electron gas (2DEG) proximitized by epitaxial aluminum.
Selective etching of the Al defines the multiterminal DC SQUID \cite{Moehle_2021}, see Figs.~\subref{fig:4termjj_device}{(a-c)}.
A schematic depiction of the circuit is given in Fig.~\subref{fig:4termjj_device}{(d)}.
The SQUIDs are designed such that two pairs of terminals in the 4TJJ each form one junction of a DC SQUID (labeled L or R).
For each SQUID, the other roughly \SI{3}{\micro\meter}-wide reference junction (labeled JJL or JJR) is much larger and therefore has a much higher critical current $I_\mathrm{c,L}^{\mathrm{ref}}$ or $I_\mathrm{c,R}^{\mathrm{ref}}$.
In this DC SQUID configuration with asymmetric critical currents, the CPR of each pair of the 4TJJ terminals can be directly measured if the SQUID loop inductance is negligible \cite{Miyazaki_2006,Della_Rocca_2007}.

\begin{figure}[t]
  \centering
  \includegraphics{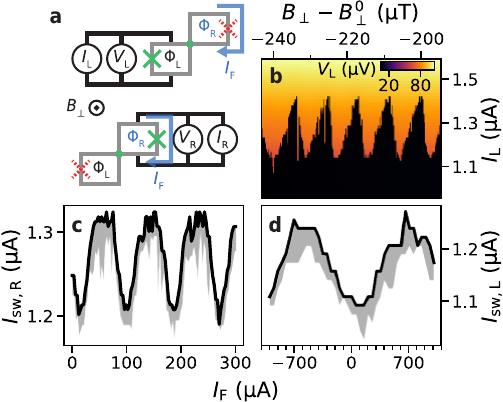}
  \caption{
    Characterization of individual SQUIDs.
    \textbf{(a)} Measurement circuit configurations for CFR measurements of the individual SQUIDs L \textbf{(b,c)},and R \textbf{(d)}, taken with the opposite SQUID's reference junction pinched-off by setting $V_\mathrm{L/R,ref}$ to a strongly negative voltage, and with current applied across one SQUID's leads with the other's floating.
    \textbf{(b)} A SQUID oscillation measurement with a single current trace per $B_\perp$ value for SQUID L.
    Current is swept positively from \SI{0}{\ampere}.
    \textbf{(c)} Maximum switching current $\mathrm{max}[I_\mathrm{sw,R}]$ (black) extracted from current traces of SQUID R as a function of the flux line current.
    Variation of the switching current across the repeated traces is shown in gray.
    \textbf{(d)} Analogous measurements of SQUID L as a function of flux line current, where we observe a very weak cross-coupling of the flux line to this SQUID.
    All $B_\perp$ scans are offset by $B_\perp^0\equiv\SI{-7.5}{\milli\tesla}$ determined as the $B_\perp$ value at which JJL and JJR showed a maximum critical current in Fraunhofer measurements.
    The field is fixed to $B_\perp=B_\perp^0$ when not being varied.
    Switching current offsets between \textbf{(b)} and \textbf{(d)} result from shifts in the effective zero-field point over time due to hysteresis in the system.
  }
  \label{fig:4termjj_characterization}
\end{figure}

Two Ti/Pd layers of gate electrodes were then patterned, separated by a \SI{20}{\nano\meter} thick AlOx dielectric from the 2DEG and from each other.
These include top gates over the reference junctions JJL and JJR of each SQUID applying voltages $V_\mathrm{L,ref}$ and $V_\mathrm{R,ref}$ respectively.
These allow us to pinch off conductance through these junctions and remove the corresponding SQUID's flux-dependent behavior.
Second, a large depletion gate in the first layer surrounding the 4TJJ (red in Fig.~\subref{fig:4termjj_device}{(b)}) allows us to apply a voltage $V_\mathrm{D}$ depleting carriers surrounding it and eventually within the junction itself.
This gate is kept grounded for all measurements except those in Appendix \ref{sec:4termjj_gatebehavior} where we investigate its behavior.
Metallic Ti/Pd probe contacts near the 4TJJ are kept electrically floating or grounded to not interfere with measurements \footnote{
  The probes are floating for all measurements except those of Figs.~\subref{fig:4termjj_characterization}{(d)}, \subref{fig:4termjj_anomalous}{(c)}, and \ref{fig:4termjj_gate_dependence}, where one is grounded.
  As switching current measurements, these measurements are unaffected by the grounded probes, since current favors traveling through the superconducting circuit until the current bias is large enough that the circuit switches into a resistive state.
}.
Lastly, a NbTiN flux bias line (blue in Figs.~\subref{fig:4termjj_device}{(a,d)}) with critical current \SI{1.05}{\milli\ampere} was sputtered around SQUID R to locally bias the magnetic flux $\Phi_\mathrm{R}$ penetrating it without significantly tuning the flux $\Phi_\mathrm{L}$ through SQUID L. 
All other gates (gray in Figs.~\subref{fig:4termjj_device}{(a,b)}) are kept grounded unless otherwise specified.
Combined with a global magnetic field with an almost fully out-of-plane component $B_\perp$ (calibrated in Appendix \ref{sec:4termjj_field_direction}), this allows for independent control of the magnetic flux through both SQUIDs.
For additional fabrication details, see Ref.~\cite{Prosko_2023_ArXiv}.
Measurements are conducted at the 20-\SI{70}{\milli\kelvin} base temperature of a dilution refrigerator.

Before proceeding to measurements, we remark that
the supercurrent across any two terminals $i,j\in\{1,2,3,4\}$ (labeled in Fig.~\subref{fig:4termjj_device}{e}) of the 4TJJ is a $\Phi_0$-periodic function of all phase differences $\phi_{i'j'}\equiv\phi_\mathrm{i'}-\phi_\mathrm{j'}$ of their superconducting wave functions \cite{Beenakker_1991}.
Because only the relative phase differences determine the junction's behavior, there are only $n-1$ phase degrees of freedom for any $n$-terminal JJ.
As we embed the 4TJJ in two DC SQUIDs, we have control of only $\phi_{12}$ and $\phi_{34}$, meaning that a third independent phase difference (\emph{eg.} $\phi_{13}$) is not directly controlled by experimentally tunable parameters.

To probe any uniquely multiterminal Josephson effects in this 4TJJ, we deduce phase shifts in its CPR through SQUID oscillation measurements, that is, measurements of the CFR.
For two-terminal JJs, an established technique for measuring their CPR is to embed the JJ in a DC SQUID containing another `reference' JJ and measure the SQUID's critical current as a function of flux.
When the reference JJ's critical current is much higher than that of the probed junction the CFR of the SQUID becomes equivalent to the CPR of the probed junction~\cite{Miyazaki_2006,Della_Rocca_2007,Nanda_2017}.
It is important to note however that in the presence of a small but finite loop inductance, this is not strictly true.
Nonetheless, we show that the CFR possesses key properties of the true CPR: its periodicity in flux, and shifts in its phase offset (see Appendix \ref{sec:4termjj_cpr_theory}).

To begin we characterize each SQUID individually, with results summarized in Fig.~\ref{fig:4termjj_characterization}.
For these measurements, one SQUID is probed while the other's leads are kept floating, see Fig.~\ref{fig:4termjj_characterization}{(a)}.
To exclude effects from the opposing SQUID, we set $V_\mathrm{L,ref}=\SI{-1}{\volt}$ or $V_\mathrm{R,ref}=\SI{-1.2}{\volt}$ to eliminate conductance through the opposite SQUID's reference junction.
This prevents applied fields from tuning the phase across the opposite SQUID's 4TJJ terminals.
The junctions in these devices have large enough self-capacitances that they are underdamped, potentially from capacitances to the nearby floating 4TJJ terminals.
This is signified by $I_{\mathrm{sw,L/R}}$ varying stochastically between values less than or equal to their critical currents $I_\mathrm{c,L/R}$
\cite{Tinkham,Haxell_2023_switch}.
An example CFR measurement is in Fig.~\subref{fig:4termjj_characterization}{(b)}, where we show switching current oscillations of SQUID L measured with a single current trace upwards from $I_\mathrm{L}=0$ for each $B_\perp$ value.
At several $B_\perp$ values $I_\mathrm{sw,L}$ appears much lower than the overall nearly-sinusoidal trend.
We observe that SQUID L exhibits a $\Phi_0$ periodicity of \SI{8.4}{\micro\tesla} as a function of $B_\perp$, expected to be roughly the same for SQUID R as they have identical lithographically-defined loop areas.
Since $I_\mathrm{sw,L/R}\leq I_\mathrm{C,L/R}$ and $I_\mathrm{sw,L/R}$ varies randomly between each trace, we focus on the maximum observed $I_\mathrm{sw,L/R}$ across repeated sweeps.
Traces are taken in a positive direction from \SI{0}{\ampere} to ensure the switching current and not retrapping current is measured.

The resulting CFR measurements of SQUIDs L and R as a function of $I_\mathrm{F}$ are plotted in Figs.~\subref{fig:4termjj_characterization}{(c)} and \subref{fig:4termjj_characterization}{(d)}.
SQUID R has a periodicity of roughly one flux quantum per \SI{85}{\micro\ampere} change in $I_\mathrm{F}$ while SQUID L has a periodicity of \SI{1.3}{\milli\ampere}, indicating that the flux-bias line almost exclusively tunes the magnetic flux through SQUID R.

\section{Flux-Tunable Nonlocal Josephson Effect}
\label{sec:4termjj_anomalouseffect}

\begin{figure*}[ht!]
  \centering
  \includegraphics{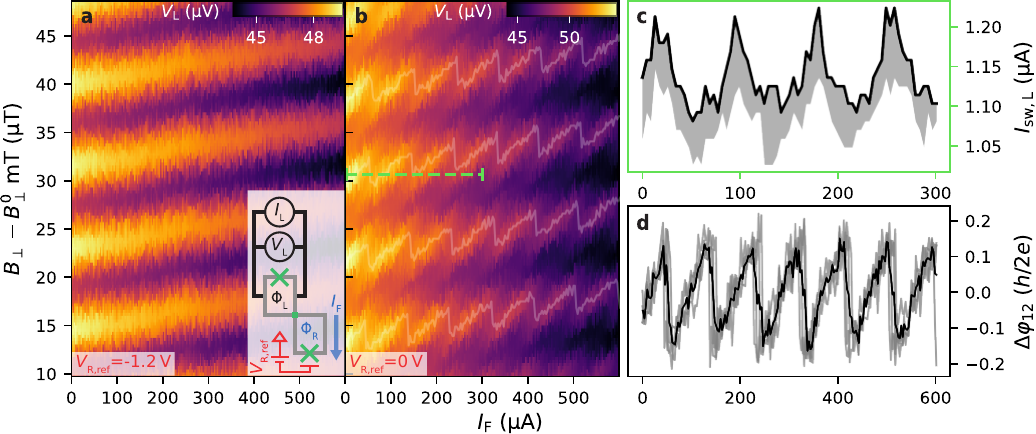}
  \caption{
    Flux-Tunable Josephson effect in Device A SQUID L.
    \textbf{(a,b)} Voltage measured with the junction in a resistive state after applying a $I_\mathrm{L}=\SI{1.1}{\micro\ampere}$ current across SQUID L with SQUID R leads floating (inset) and \textbf{(a)} JJR pinched-off or \textbf{(b)} with it at \SI{0}{\volt}.
    Extracted $V_\mathrm{L}$ maxima are plotted in faint white, and show a zig-zag like variation with a periodicity equal to SQUID R's $h/2e$-periodicity in flux line current.
    \textbf{(c)} Extracted maximum switching current (black) along the green line in \textbf{(b)}, calculated from 10 current traces swept upwards from zero current.
    Variation across the repeated current traces is shown in gray.
    Maxima in $I_\mathrm{sw,L}$ correspond to maxima in the measured voltage of \textbf{(b)}.
    Cross coupling of the flux-line field to SQUID L is negligible in this flux-current range, so the observed oscillations in the CPR are purely due to the nonlocal superconducting phase difference across terminals 3 and 4.
    \textbf{(d)} Phase offsets of SQUID L's local CPR (gray) for the four flux periods visible in \textbf{(b)}, and their mean (black), calculated from \textbf{(b)} up to a constant offset.
    We observe a $\phi_0$ tunability in a range greater than 0.2 flux quanta.
  }
  \label{fig:4termjj_anomalous}
\end{figure*}

We proceed by conducting measurements involving both SQUIDs, aiming to probe nonlocal effects of the phase difference across two terminals of the 4TJJ on the CPR through the other two.
To do so, we float the leads of SQUID R but keep JJR conducting unless otherwise specified, measuring the voltage $V_\mathrm{L}$ across SQUID L as a function of $B_\perp$ and $I_\mathrm{F}$.
Recall that $B_\perp$ roughly equally tunes $\Phi_\mathrm{L}$ and $\Phi_\mathrm{R}$ while $I_\mathrm{F}$ almost exclusively tunes $\Phi_\mathrm{R}$ (see Figs.~\subref{fig:4termjj_characterization}{(c,d)}).
This means we can fully navigate the space of phase differences $\phi_{12}$ and $\phi_{34}$ by sweeping these two parameters.
Our results are summarized in Fig.~\ref{fig:4termjj_anomalous}.

In Figs.~\subref{fig:4termjj_anomalous}{(a,b)} we fix $I_\mathrm{L}=\SI{1.1}{\micro\ampere}$ to a value near the SQUID L critical current and measure $V_\mathrm{L}$.
The SQUID voltage in its resistive state is a function of its critical current, and so must have the same periodicity and phase offset \cite{Tinkham}.
Hence, from the positions of extrema in $V_\mathrm{L}$ we can extract the relative value of the $\phi_0$ offset $\Delta\phi_{12}$ across terminals 1 and 2.
When JJR is closed as in Fig.~\subref{fig:4termjj_anomalous}{(a)} so that $\phi_{34}$ is not tuned by $B_\perp$ or $I_\mathrm{F}$, only local SQUID oscillations as a function of $B_\perp$ are visible, tilted upwards due to the small cross coupling of $I_\mathrm{F}$ into $\Phi_\mathrm{L}$.
Remarkably however, when $V_\mathrm{R,ref}=\SI{0}{\volt}$ as in Fig.~\subref{fig:4termjj_anomalous}{(b)}, lobes of SQUID oscillation maxima appear in a zig-zag pattern.
The lines of maximum $V_\mathrm{L}$ (highlighted in gray) are oriented diagonally along the $B_\perp$ and $I_\mathrm{F}$ axes in a different direction than the maxima lines in Fig.~\subref{fig:4termjj_anomalous}{(a)}.
This feature thus arises due to a variation in $\Phi_\mathrm{R}$ changing $\phi_{34}$.
In other words, the supercurrent through terminals 1 and 2 of the 4TJJ is tuned by the phase difference across two completely different terminals of the junction, manifesting a a flux-tunable $\phi_0$-junction.
Three-terminal circuits of two JJs sharing a common lead have enabled observations of similar nonlocal Josephson effects \cite{Matsuo_2022,Haxell_2023,Matsuo_2023_ArXiv_nonlocal} as well as one controlled by excess spins in the junction \cite{Strambini_2020}.
In these experiments, nonlocal coupling between junctions was claimed to arise due to the direct wave function overlap of ABSs in either junction.
While these experiments have no obvious analog in four-terminal circuits, the effect observed here may be described as the limit of ABS wave functions in two JJs completely merging together \cite{Coraiola_2023}.
We compare these experiments further in Sec.~\ref{sec:4termjj_2tjj_arrays}.

To make this observation concrete, in Fig.~\subref{fig:4termjj_anomalous}{(c)} we measure repeated current traces of SQUID L along the green line highlighted in Fig.~\subref{fig:4termjj_anomalous}{(b)}.
The maximum observed switching current across all traces is shown as a black line while the variation of $I_\mathrm{sw,L}$ is shown in gray.
Large $\gtrsim$\SI{0.1}{\micro\ampere} oscillations in the switching currents are visible as a function of $I_\mathrm{F}$ with the periodicity of SQUID oscillations in SQUID R.
Since a change of over \SI{1}{\milli\ampere} in $I_\mathrm{F}$ is required to tune $\Phi_\mathrm{L}$ by one flux quantum due to direct cross coupling, these SQUID oscillations are purely due to coupling of $\phi_{34}$ to the CPR between terminals 1 and 2.
Additionally, we quantify the degree to which the $\phi_0$ offset between terminals 1 and 2 can be tuned in Fig.~\subref{fig:4termjj_anomalous}{(d)}.
From the positions of successive SQUID oscillation maxima plotted in Fig.~\subref{fig:4termjj_anomalous}{(b)} in white, we extract the relative change $\Delta\phi_{12}$ of each peak as a function of $I_\mathrm{F}$.
Since we can only extract the relative apparent $\phi_{12}$ offset of the junction, we arbitrarily define $\Delta\phi_{12}=0$ as the maximum position at $I_\mathrm{F}=0$.
A linear offset is also subtracted from each maximum to remove the effect of cross coupling between $I_\mathrm{F}$ and $\Phi_\mathrm{L}$, and the result is converted into units of flux quanta from the $B_\perp$-periodicity of SQUID L oscillations.
The average $\Delta\phi_{12}$ across all maxima is plotted in black, while individual peak oscillations are in gray.
We see that the $\phi_0$-junction formed across terminals 1 and 2 can have its phase offset tuned continuously in a range of over $0.2\Phi_0$.
Behavior consistent with this is observed for analogous measurements of SQUID R, shown in Appendix \ref{sec:4termjj_squid_r}, and similar effects are also seen in measurements of a second device of identical design in Appendix \ref{sec:4termjj_second_device}.

Importantly, the $\Phi_0$-periodic circulating current in SQUID R could couple trivially to the flux through SQUID L via the loops' mutual inductance in the absence of any four-terminal junction effects.
Due to the device design maximizing the spatial separation between SQUID loops however (Fig.~\subref{fig:4termjj_device}{(a)}), such a coupling could not cause the observed strength of oscillations in $\Delta\phi_{12}$: the magnetic field produced by such loops carrying currents of less than \SI{2}{\micro\ampere} produces less than \SI{1}{\percent} of a flux quantum in the opposing loop.

\section{Multiterminal Junctions as Two-Terminal Junction Arrays}
\label{sec:4termjj_2tjj_arrays}

\begin{figure*}[t]
  \centering
  \includegraphics{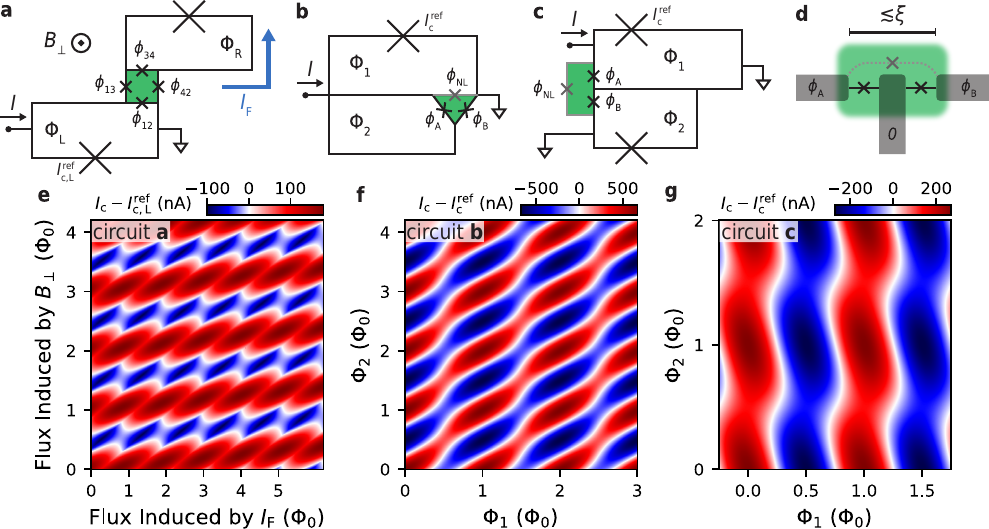}
  \caption{
    Simulations of multiterminal JJs as networks of two-terminal JJs.
    \textbf{(a-c)} Two-terminal JJ models of the 4TJJ in this article \textbf{(a)}, and of the Andreev molecule device geometries of Ref.~\cite{Haxell_2023} \textbf{(b)} and Ref.~\cite{Matsuo_2023_ArXiv_nonlocal} \textbf{(c)}.
    We model the 4TJJ (green) as four two-terminal JJs between each neighboring terminal.
    For the Andreev molecule devices we model two nearby JJs (labeled A and B with phase differences $\phi_\mathrm{A}$ and $\phi_\mathrm{B}$) as well as an incidental JJ (gray, with phase difference $\phi_\mathrm{NL}$) shunting the two JJs through the same semiconducting region (green).
    Large crosses indicate reference JJs with critical currents much larger than those of the JJs in the green regions.
    \textbf{(d)} Close-up schematic of the layout of leads (gray) in the green regions of \textbf{(c)} and \textbf{(d)} patterned over a semiconducting 2DEG (green).
    \textbf{(e)} Simulated critical current $I_\mathrm{c}$ of SQUID L with SQUID R floating as a function of $B_\perp$ and $I_\mathrm{F}$ in units of flux quanta, modeling the 4TJJ--SQUID circuit with the circuit model in \textbf{(a)}.
    We use $I_\mathrm{c,12}=I_\mathrm{c,34}=\SI{100}{\nano\ampere}$ and $I_\mathrm{c,13}=I_\mathrm{c,24}=\SI{80}{\nano\ampere}$.
    Critical current maxima qualitatively reproduce the zig-zag pattern observed in Fig.~\subref{fig:4termjj_anomalous}{(b)}.
    \textbf{(f,g)} Simulated critical current of the circuits in \textbf{(b)} and \textbf{(c)} as a function of the fluxes $\Phi_{1}$ and $\Phi_{2}$ threading the loops.
    We use $I_\mathrm{c,A/B}=\SI{450}{\nano\ampere}$ \textbf{(f)} and $I_\mathrm{c,A/B}=\SI{200}{\nano\ampere}$ \textbf{(g)} as in Refs.~\cite{Haxell_2023} and \cite{Matsuo_2023_ArXiv_nonlocal}, respectively.
    Additionally, we choose $I_\mathrm{c,NL}=\SI{180}{\nano\ampere}$ \textbf{(e)} and $I_\mathrm{c,NL}=\SI{70}{\nano\ampere}$.
    In all plots, $I_\mathrm{c}$ is offset by the reference junction's critical current.
  }
  \label{fig:4termjj_2tjj_array}
\end{figure*}

In multiterminal JJs and Andreev molecule devices, tunable $\phi_0$-junctions are often considered a signature of behavior distinct to hybridized ABSs \cite{Pillet_2019,Kocsis_2023_ArXiv} or of an ABS spectrum distinctly associated with multiterminal JJs \cite{Omelyanchouk_2000}.
In this section, we demonstrate that modeling these systems with networks of two-terminal JJs produces a tunable $\phi_0$-junction behavior which may be difficult to distinguish from the case where bound states in these junctions are truly hybridized into a multiterminal JJ or an Andreev molecule.
Namely, while several superconducting terminals connected to a semiconducting region smaller than the superconducting coherence length $\xi$ is naturally described by a hybridized ABS spectrum \cite{Beenakker_1991}, similar nontrivial behavior in the CPR is also expected for a network of two-terminal JJs connecting each terminal.
As a minimal model, we consider circuits of JJs neglecting linear inductances and capacitive effects, and approximate each junction as possessing a sinusoidal CPR.
As examples, we consider a two-terminal JJ network designed to emulate the results measured in Fig.~\subref{fig:4termjj_anomalous}{(b)} and others to reproduce results of recent experiments observing CPRs consistent with Andreev molecule effects \cite{Haxell_2023,Matsuo_2023_ArXiv_nonlocal}.

Beginning with the 4TJJ embedded in two asymmetric DC SQUIDs as in this experiment, we model the 4TJJ as four two-terminal JJs of critical current $I_{\mathrm{c},ij}$ coupling each neighboring superconducting terminal $i,j\in\{1,2,3,4\}$, see Fig.~\subref{fig:4termjj_2tjj_array}{(a)}.
From switching current measurements of SQUID L with its reference junction pinched off, we estimate the equivalent critical current between terminals 1 and 2 as $I_\mathrm{c,12}=I_\mathrm{c,34}=\SI{100}{\nano\ampere}$, approximating the same between terminals 3 and 4 by symmetry in the device design.
Since the distance between terminals 1 and 3 or 2 and 4 is larger, we select $I_\mathrm{c,13}=I_\mathrm{c,24}=\SI{80}{\nano\ampere}$ to approximate the qualitative behavior of the CFR maxima seen in Fig.~\subref{fig:4termjj_anomalous}{(b)}.
By applying Kirchhoff's current law and flux quantization while assuming the flux threading the 4TJJ (green in Fig.~\subref{fig:4termjj_2tjj_array}{(a)}) is negligible, we can calculate the critical current across SQUID L \cite{Rasmussen_2021}.
Results are shown in Fig.~\subref{fig:4termjj_2tjj_array}{(e)}.
While the functional dependence of $V_\mathrm{L}$ measured at fixed current in the resistive state is not expected to precisely match the CPR, the simulated critical current exhibits a similar zig-zag pattern in the positions of maximum critical current.
This indicates that a tunable $\phi_0$-junction alone is not unique to multiterminal JJ behavior.
Namely, this demonstrates that while the lithographic design of the devices measured here contain a 4TJJ, the $\phi_0$-junction could appear even if the ABSs formed between each pair of terminals were not hybridized with any other ABSs.

As further examples, we model devices expected to host ABSs hybridized into Andreev molecules \cite{Pillet_2019,Kocsis_2023_ArXiv}.
These devices consist of two two-terminal JJs of critical current $I_\mathrm{c,A/B}$ and phase difference $\phi_\mathrm{A/B}$ sharing a common superconducting lead and separated by a distance on the order of the $\xi$, depicted in Fig.~\subref{fig:4termjj_2tjj_array}{(d)}.
Due to wave function overlap between the ABS in each junction, an Andreev molecular state is expected to form.
A stark signature of this state is a phase offset in one junction tunable by the phase difference across the other junction \cite{Pillet_2019,Kocsis_2023_ArXiv}.
In practice however, measured CPRs of Andreev molecule devices have formed the two JJs from a common region of semiconducting material \cite{Haxell_2023,Matsuo_2023_ArXiv_nonlocal}.
Since their separation is less than the superconducting coherence length, supercurrent could pass between the outer terminals in the absence of any hybridization of the ABSs in the intended junctions.

We thus model the device geometries of Refs.~\cite{Haxell_2023} and \cite{Matsuo_2023_ArXiv_nonlocal} with the circuits shown in Figs.~\subref{fig:4termjj_2tjj_array}{(b)} and \subref{fig:4termjj_2tjj_array}{(c)} respectively.
Nonlocal effects are modeled by a JJ directly coupling the outer leads with phase difference $\phi_\mathrm{NL}$ and critical current $I_\mathrm{c,NL}<I_\mathrm{c,A/B}$.
Since each pair of leads is expected to support supercurrent in the absence of the remaining lead, this is roughly equivalent to considering the three-terminal junction with hybridization of the ABSs formed in each junction neglected.
Specific values of $I_\mathrm{c,A/B}$ are extracted directly from Refs.~\cite{Haxell_2023,Matsuo_2023_ArXiv_nonlocal}, while  $I_\mathrm{c,NL}$ values are chosen to best match their measurements.
The resulting simulations are shown in Figs.~\subref{fig:4termjj_2tjj_array}{(f)} and \subref{fig:4termjj_2tjj_array}{(g)}.
They bear remarkable similarity with the measurements, in particular producing similar $\phi_0$-junction tunability to these experiments in the absence of any ABS hybridization.
For more details of these calculations, see Appendix \ref{sec:4termjj_2tjj_models}.

Naturally, as these junctions are defined in a region smaller than $\xi$, hybridization between ABSs in the junctions is expected to have contributed to the measured CPRs \cite{Kornich_2020}.
The above modeling shows, however, that the level of ABS hybridization in existing experiments does not yield CPRs easily distinguishable from the case of a non-interacting three-terminal junction.
To exclude this trivial coupling between leads in Andreev molecule devices, measuring similar devices designed with no direct path through the semiconductor between the outer superconducting leads rules out this shunting effect.
For example, each junction could be formed from different semiconducting nanowires \cite{Kurtossy_2021,Matsuo_2022}.
Importantly, tunneling spectroscopy measurements of the semiconducting region could also reveal an ABS spectrum exhibiting anticrossings indicative of hybridization between the ABSs \cite{Pillet_2019,Coraiola_2023,Coraiola_2023_ArXiv_spin}.

\section{Conclusions \& Outlook}
\label{sec:4termjj_outlook}

We have studied a 4TJJ by embedding it in two asymmetric DC SQUIDs, observing nontrivial properties of the CPR of a 4TJJ.
Namely, we were able to measure SQUID oscillations of two pairs of terminals forming the junction and independently tune two of the three independent phase differences controlling it.
From these measurements, we observed a nonlocal Josephson effect: two terminals of the 4TJJ behaved as a $\phi_0$-junction with a phase offset tunable by the nonlocal flux biasing the phase difference across two independent junction terminals.
This tunability had a range exceeding $0.2\Phi_0$, and allows the 4TJJ to serve as a superconducting current switch \cite{Alidoust_2012}.
Modeling multiterminal junctions as two-terminal JJ arrays, we also found that $\phi_0$-junction effects alone are not sufficient evidence of hybridization between ABSs in the junction.

Future devices with a barrier gate separating the lead pairs could demonstrate for the first time tunable direct wave function overlap between phase-tunable ABSs.
Coupling between relatively distant ABSs mediated by supercurrents or photons in a macroscopic circuit has been observed \cite{PitaVidal_2023_ArXiv,Cheung_2023_ArXiv}, but demonstrating a local coupling would
enable the formation of Andreev molecule-based quasiparticle charge qubits \cite{Pillet_2019,Geier_2023_ArXiv}, or densely-spaced conventional Andreev qubits.
Andreev molecule devices where ABSs are coupled with a superconducting lead in between have exhibited hybridization effects \cite{Coraiola_2023,Matsuo_2023_ArXiv_molecule}, but their coupling is fixed by the superconducting lead dimensions \cite{Pillet_2019,Kornich_2020}.
A tunable wave function overlap with directly tunnel-coupled ABSs in JJs provides an alternate mechanism for realizing qubits based on ABSs or Kitaev chains \cite{Fulga_2013}, allowing for readout via inductive coupling of resonators to the phase-biased loops containing each JJ \cite{Zazunov_2003,Janvier_2015}.
Last and most notably, with control over one more phase difference in the 4TJJ, Weyl singularities in this system's subgap state spectrum could be probed \cite{Riwar_2016,Xie_2018,Barakov_2023}.

\section*{Data Availability}

Raw data and scripts for plotting the figures in this publication are available from Zenodo \cite{datarepo_4termjj}.

\begin{acknowledgements}
  The authors would like to thank O.W.B. Benningshof and J.D. Mensingh for technical assistance with the cryogenic electronics.
  We would also like to thank M. Coraiola and F. Nichele for input on the manuscript, and A.R. Akhmerov and K. Vilkelis for discussions on theoretical modeling.
  The authors also acknowledge financial support from Microsoft Quantum and the Dutch Research Council (NWO).
\end{acknowledgements}

\section*{Author Contributions}
The devices were designed by W.D.H., fabricated by W.D.H. and I.K. using a 2DEG heterostructure provided by D.X., C.T., and M.J.M..
Data was measured, analyzed, and modeled by C.G.P..
I.K. assisted with designing the measurement setup, and S.G. supervised the project.
C.G.P. wrote the manuscript with input from all authors.

\begin{appendix}


  \section{Gate Performance}
  \label{sec:4termjj_gatebehavior}

\begin{figure}[t]
    \centering
    \includegraphics{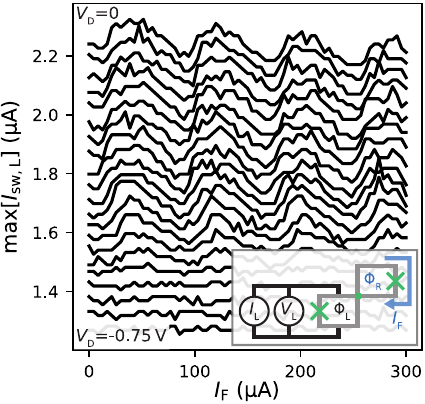}
    \caption{
      Depletion Gate Dependence of nonlocal SQUID L oscillations in Device A.
      As the depletion gate $V_\mathrm{D}$ is swept, five current traces at each $I_\mathrm{F}$ are measured and used to extract $\mathrm{max}[I_\mathrm{sw,L}]$.
      Traces are offset from each other by \SI{50}{\nano\ampere} for clarity.
    }
    \label{fig:4termjj_gate_dependence}
  \end{figure}

  In Fig.~\ref{fig:4termjj_gate_dependence} we plot a characterization of SQUID L oscillations as a function of the nonlocal flux $\Phi_\mathrm{R}$ tuned by $I_\mathrm{F}$ in Device A as the central device depletion gate voltage (red in Fig.~\subref{fig:4termjj_device}{(b)}) $V_\mathrm{D}$ is swept down from \SI{0}{\volt}.
  Repeating multiple current traces at each $I_\mathrm{F}$ value, we plot the maximum observed switching current as it is closest to the SQUID critical current.
  As the gate depletes carriers in the 4TJJ, the amplitude of oscillations decreases until none are observed by $V_\mathrm{D}=-\SI{0.75}{\volt}$.
  When $V_\mathrm{D}=0$, the SQUID oscillations are slightly skewed to the left, and this skewness also appears to reduce, leaving the oscillations more sinusoidal at intermediate $V_\mathrm{D}$ values.
  A detailed investigation of the influence of patterned gates on the 4TJJ characteristics was made impossible by the misalignment of gates in the second layer (gray in Fig.~\subref{fig:4termjj_device}).
  These gates were designed to tune the chemical potential selectively between pairs of terminals, enable tunneling spectroscopy with the normal metal probes, and tunably isolate SQUID L from SQUID R.

  \section{Field Direction Calibration}
  \label{sec:4termjj_field_direction}

  \begin{figure}[hbt!]
    \centering
    \includegraphics{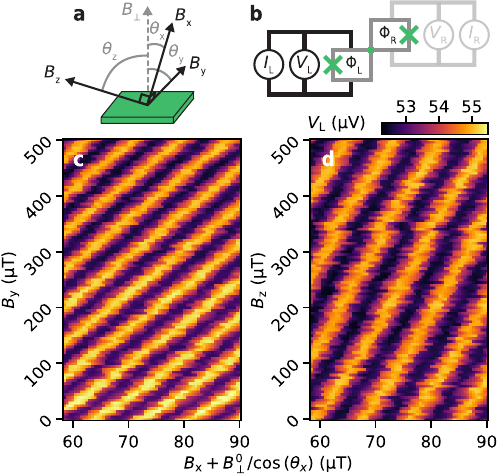}
    \caption{
      Field direction calibration in Device B.
      \textbf{(a)} Diagram schematically depicting how the applied field in the $B_\mathrm{x}$ direction translates to an out-of-plane component $B_\perp$ as well as small components $B_\mathrm{y}$ and $B_\mathrm{z}$ in the $y$- and $z$-directions respectively.
      The angle $\theta_\mathrm{x}$ is exaggerated for clarity.
      \textbf{(b)} Measurement configuration for field orientation calibration.
      \textbf{(c),(d)} SQUID L oscillations as a function of $B_\mathrm{x}$ and $B_\mathrm{y}$ \textbf{(c)} and $B_\mathrm{z}$ \textbf{(d)} with a fixed current $I_\mathrm{L}=I_\mathrm{R}=\SI{1.45}{\micro\ampere}$ applied by independent sources across both SQUIDs, though these measurements are only used to infer the oscillation periodicity of SQUID L in each field direction.
      For these measurements, $V_\mathrm{D}=\SI{-0.15}{\volt}$ while each second-layer gate except the rightmost one (gray gates in Fig.~\subref{fig:4termjj_device}{(b)}) has \SI{+0.8}{\volt} applied.
      From these scans, we infer a mean peak spacing of oscillations along each axis of approximately $\Delta B_\mathrm{x}=\SI{8.5}{\micro\tesla}$, $\Delta B_\mathrm{y}=\SI{64}{\micro\tesla}$, and $\Delta B_\mathrm{z}=\SI{0.14}{\milli\tesla}$.
      Measurements are with respect to an approximate zero-field point along the $B_\mathrm{x}$ direction of $B_\perp^0=\SI{-7.15}{\milli\tesla}$ calculated from Fraunhofer pattern measurements of the reference junctions.
    }
    \label{fig:4termjj_field_orientation}
  \end{figure}

  For all measurements in the main text of this manuscript, the external magnetic field used $B_\mathrm{x}$ (as opposed to the field generated by the flux current $I_\mathrm{F}$) was along a three-axis magnet's `x' direction, mostly out-of-plane of the chip, see Fig.~\subref{fig:4termjj_field_orientation}{(a)}.
  To calculate the out-of-plane component $B_\perp$ as labeled in both figures, we calibrate the field direction with measurements on Device B, summarized in Fig.~\ref{fig:4termjj_field_orientation}.
  Measuring SQUID oscillations of SQUID L in its resistive state, akin to the measurements of Figs.~\subref{fig:4termjj_anomalous}{(a)} and \subref{fig:4termjj_anomalous}{(b)}, we extract the periodicity of oscillations along each of the magnets three axes.
  From these measurements, we infer that the angles $\theta_j$ for $j\in\{\mathrm{x},\mathrm{y},\mathrm{z}\}$ of field $B_j$ with respect to the out-of-plane vector are $\theta_\mathrm{x}=\SI{8.3}{\degree}$, $\theta_\mathrm{y}=\SI{82}{\degree}$, and $\theta_\mathrm{z}=\SI{86}{\degree}$.
  This implies that the $h/2e$ periodicity of SQUID L with respect to $B_\perp$ is approximately \SI{8.4}{\micro\tesla} in Device B, consistent with the devices' loop areas (see Fig.~\subref{fig:4termjj_device}{(a)}).

  As Device A may have been loaded in a different direction with respect to the magnet compared to Device B, these angles are not the same for Device A.
  Despite this, because Device A has the same lithographical design as Device B, its SQUIDs' oscillation periodicities are expected to be the same.
  Hence, from the $B_\mathrm{x}$ periodicity of SQUID L extracted from the data of Fig.~\subref{fig:4termjj_anomalous}{(a)} before converting the $B_\mathrm{x}$ axis to $B_\perp$, we infer $\theta_\mathrm{x}=\SI{14}{\degree}$ for Device A.
  This enables us to calculate $B_\perp$ from the applied field.

  \section{Current-Phase Relations of Four-Terminal Junctions embedded in Asymmetric SQUIDs}
  \label{sec:4termjj_cpr_theory}

  On its own, it is impossible to measure the CPR of the 4TJJ because the phase differences $\phi_{ij}$ across its terminals $\{1,2,3,4\}$ cannot be controlled.
  Embedding each pair of terminals (namely $\{1,2\}$ and $\{3,4\}$) from the 4TJJ into a DC SQUID penetrated by magnetic fluxes $\Phi_\mathrm{L}$ and $\Phi_\mathrm{R}$ allows control of $\phi_{12}$ and $\phi_{34}$, respectively, through tuning of these fluxes.
  Reference junctions JJL and JJR must also be embedded in each SQUID loop to prevent the SQUID's critical current from being too large to practically measure.
  In that case, the supercurrent across SQUID L when the other SQUID's leads are floating is
  \begin{equation}
    I = I_{\mathrm{c,L}}^\mathrm{ref}f_\mathrm{L}(\phi_\mathrm{L}) + I_{\mathrm{c,M}}f_{12}(\phi_{12},\phi_{34},\phi_{13})
  \end{equation}
  where $\phi_{\mathrm{L/R}}$ is the phase difference across reference junction L/R with critical current $I_\mathrm{c,L/R}^\mathrm{ref}$, $I_{\mathrm{c,M}}$ is the critical current of the 4TJJ, and $f_\mathrm{L}$ and $f_{12}$ are some $\Phi_0$-periodic functions such that $\vert f_\mathrm{L}\vert,\vert f_{12}\vert \leq 1$ \cite{Golubov_2004}.
  In other words, $I_{\mathrm{cM}}f_{ij}$ is the current phase relation of the 4TJJ between leads $i$ and $j$, which depends on all phase differences across it.

  We assume the other SQUID's leads are floating so that current into the 4TJJ is conserved.
  By flux quantization, we have that
  \begin{equation}\label{eq:4termjj_flux_quantization}
    \phi_{12} - \phi_\mathrm{L}
    = \frac{2\pi(\Phi_\mathrm{L} + L_\mathrm{L}J_\mathrm{L})}{\Phi_0}
    \mod{(2\pi)}
  \end{equation}
  where $L_\mathrm{L/R}$ and $J_\mathrm{L/R}$ are the self-inductance and the circulating current around SQUID L/R.
  Note that $J_\mathrm{L/R}$ must itself be $\Phi_0$ periodic in flux.
  Because we have that $I_{\mathrm{c,L/R}}^\mathrm{ref}\gg I_{\mathrm{c,M}}$, the phase difference $\phi_\mathrm{L}$ will adjust itself to whichever value $\phi_\mathrm{L}^{\mathrm{max}}$ maximizes the current flowing through JJL, since this in turn maximizes the SQUID's critical current.
  As the 4TJJ by comparison has a negligible effect on the SQUID supercurrent, the phase difference $\phi_{12}$ adjusts to the value allowing the flux quantization condition to be satisfied: $\phi_{12} = 2\pi(\Phi_\mathrm{L}+L_\mathrm{L}J_\mathrm{L})/\Phi_0 + \phi_\mathrm{L}^{\mathrm{max}}$.
  Meanwhile, the opposite SQUID has no current bias applied directly across it, and before the circuit critical current is reached, cannot have more than $I_\mathrm{c,M}$ circulating through it.
  Because $I_\mathrm{c,R}\gg I_\mathrm{c,M}$, this means  $\phi_\mathrm{R}$ must be at a value corresponding to a near zero fraction of its critical current, namely $\phi_\mathrm{R}\approx 0$.
  The critical current of the entire SQUID is then:
  \begin{widetext}
    \begin{equation}
      I_{\mathrm{c,L}} = I_\mathrm{c,L}^\mathrm{ref} + I_{\mathrm{c,M}}f_{12}\left(\frac{2\pi(\Phi_\mathrm{L}+L_\mathrm{L}J_\mathrm{L})}{\Phi_0} + \phi_\mathrm{L}^{\mathrm{max}},\frac{2\pi(\Phi_{R}+L_\mathrm{R}J_\mathrm{R})}{\Phi_0},\phi_{13}\right).
    \end{equation}
  \end{widetext}
  When the loop inductances $L_\mathrm{L/R}$ are negligible, note that $\phi_{12}$ and $\phi_{34}$ are linear in the applied flux, so we can directly control the phase difference across each pair of the 4TJJ's leads by tuning $\Phi_\mathrm{L/R}$.
  Hence, the critical current of the SQUID is equal to the CPR of the 4TJJ across two terminals shifted by the critical current of the reference junction and skewed by non-zero loop inductances $L_\mathrm{L}$ and $L_\mathrm{R}$.

  Summarily, we have that $\phi_{12/34}=2\pi(\Phi_\mathrm{L/R}+L_\mathrm{L/R}J_\mathrm{L/R})/\Phi_0$ plus a constant offset.
  Conservatively estimating that the individual SQUIDs have inductances of $L_\mathrm{L/R}<\SI{100}{\pico\henry}$ and circulating currents bounded by $J_\mathrm{L/R} \leq (I_\mathrm{c,L/R}^\mathrm{ref}+I_\mathrm{c,M})/2\approx\SI{0.7}{\micro\ampere}$, circulating currents perturb $\phi_{12/34}$ by less than $0.09$ radians.
  Without knowing $L_\mathrm{L/R}$ precisely, the CFR still possesses key properties of the true CPR due to the periodicity of $J_\mathrm{L/R}$: its periodicity in flux, and shifts in its phase offset.

  \section{Nonlocal Flux Dependence in \mbox{SQUID R}}
  \label{sec:4termjj_squid_r}

  \begin{figure}[ht!]
    \centering
    \includegraphics{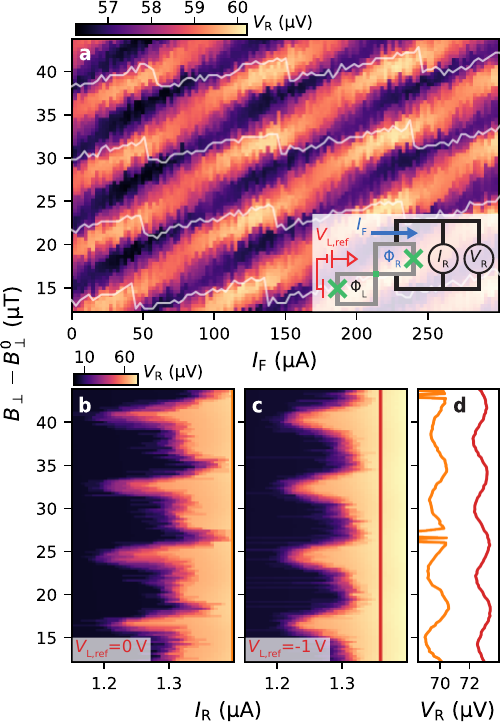}
    \caption{
      Nonlocal Josephson Effect SQUID R on Device A.
      \textbf{(a)} Voltage $V_\mathrm{R}$ measurements at a fixed current $I_\mathrm{R}=\SI{1.25}{\micro\ampere}$ across SQUID R with SQUID L leads floating (inset).
      Because $B_{\perp}$ tunes both the nonlocal flux $\Phi_\mathrm{L}$ and local flux $\Phi_\mathrm{R}$, SQUID L oscillations appear perpendicular to the diagonal.
      Nonlocal effects from SQUID L perturb the path of local SQUID oscillations in a zig-zag fashion, and cause oscillations along the diagonal of the voltage measured.
      In white, positions of SQUID L oscillation maxima from Fig.~\subref{fig:4termjj_anomalous}{(b)} are plotted to emphasize these correlations.
      \textbf{(b)} Current-voltage traces of SQUID R oscillations at $I_\mathrm{F}=0$ with the SQUID L reference junction open, where strongly non-sinusoidal effects are observable, in addition to apparent minima lobes in $I_\mathrm{c,R}$ spaced by half the flux periodicity.
      Each full $B_{\perp}$ and current sweep is repeated 25 times and averaged.
      \textbf{(c)} As in \textbf{(b)}, but with the SQUID L reference junction closed off $V_\mathrm{L,ref}=-\SI{1}{\volt}$ and averaged 15 times.
      SQUID oscillations are still highly non-sinusoidal, but distinctly lack the additional lobes present in \textbf{(b)}.
      \textbf{(d)} Averaged linecuts at fixed current taken along the vertical lines in \textbf{(b)} and \textbf{(c)}.
      The positions of maxima and minima in these linecuts align with extrema in the full current sweep measurements, but the resistive voltage lacks the additional $(\Phi_0/2)$-periodic oscillation component and appears sinusoidal.
      As in the main text, $B_\perp^0=-\SI{7.5}{\milli\tesla}$ for Device A.
    }
    \label{fig:4termjj_squid_r_nonlocal}
  \end{figure}

  For comparison with Fig.~\ref{fig:4termjj_anomalous}, we measure the nonlocal coupling of $\Phi_\mathrm{L}$ into the SQUID R oscillations containing the junction formed across terminals 3 and 4, with results summarized in Fig.~\subref{fig:4termjj_squid_r_nonlocal}.
  The manifestation or strength of nonlocal effects is distorted in this case because the parameter tuning the nonlocal flux $\Phi_\mathrm{L}$ is $B_{\perp}$, which nearly equally tunes $\Phi_\mathrm{R}$.
  Conversely, in this case the parameter $I_\mathrm{F}$ tuning the local flux $\Phi_\mathrm{R}$ has a negligible effect on the nonlocal flux.
  As in Fig.~\subref{fig:4termjj_anomalous}{(b)}, we first fix the current $I_\mathrm{R}$ across SQUID R near $I_\mathrm{c,R}$ and measure voltage $V_\mathrm{R}$ in Fig.~\subref{fig:4termjj_squid_r_nonlocal}{(a)}.

  Despite $B_{\perp}$ tuning both $\Phi_\mathrm{L}$ and $\Phi_\mathrm{R}$, nonlocal features are still visible in Fig.~\subref{fig:4termjj_squid_r_nonlocal}{(a)}.
  Namely, in addition to the expected diagonal $V_\mathrm{R}$ oscillations associated with local SQUID R oscillations, the intensity of the voltage oscillations changes periodically with SQUID L's $B_{\perp}$ periodicity.
  To emphasize this, we plot the positions of SQUID L oscillation maxima extracted from Fig.~\subref{fig:4termjj_anomalous}{(b)} in white, where it aligns with the local maxima in $V_\mathrm{R}$ along the diagonal.
  Additionally, a minor zig-zag perturbation of the SQUID R oscillations from a simple diagonal path is visible, but because of the strong dependence of $\Phi_\mathrm{R}$ on both $B_{\perp}$ and $I_\mathrm{F}$, it is difficult to quantify the degree to which $\Phi_\mathrm{L}$ tunes this junction into a $\phi_0$-junction.

  To investigate the degree to which four-terminal Josephson effects are present across terminals 3 and 4, we plot full current traces with JJL open (Fig.~\subref{fig:4termjj_squid_r_nonlocal}{(b)}) and closed off (Fig.~\subref{fig:4termjj_squid_r_nonlocal}{(c)}).
  Each plot is averaged over many $I_\mathrm{R}$ and $B_{\perp}$ measurements to alleviate effects from instability of the SQUIDs as a function of $B_{\perp}$.
  Remarkably, when JJL has a finite critical current, lobes in the SQUID oscillations are visible spaced by half the SQUID L $B_{\perp}$ periodicity.
  Conversely, when JJL is closed, these higher harmonic lobes vanish, though the oscillations remain significantly non-sinusoidal.
  In Fig.~\subref{fig:4termjj_squid_r_nonlocal}{(a)} only $\Phi_0$-periodic oscillations are visible along both axes, however.
  Based on existing theories of 4TJJs \cite{Omelyanchouk_2000,Amin_2001}, this behavior is actually expected, and can be thought of as two flux quanta being threaded into the SQUIDs per $\Phi_0$ period of $B_{\perp}$.
  For such junctions, there are regions of $(\phi_{12},\phi_{34})$-space where phase slips of the JJ occur due to the appearance of vortex states inside the junction, producing additional local extrema in the CPR along lines of equal $\phi_{12}$ and $\phi_{34}$.

  Interestingly, these additional extrema in the critical current are not reflected in the voltage measured in the resistive state, emphasized by linecuts in Fig.~\subref{fig:4termjj_squid_r_nonlocal}{(d)}.
  A nearly sinusoidal resistive-state voltage is observed even when $V_\mathrm{L,ref}=0$, indicating that certain features of the CPR are not noticeably reflected in $V_\mathrm{R}$.
  From the perspective of a resistively and capacitively shunted junction (RCSJ) model \cite{Tinkham}, for example, this is possible for an underdamped junction possessing a substantial self-capacitance.
  Due to the stochastically varying switching current observed in current traces (see the averaged traces of Fig.~\ref{fig:4termjj_characterization} for example), it is clear that the 4TJJ is underdamped.
  Nonetheless, we emphasize that the periodicity and phase offset of the fixed-current measurements accurately reflect that of the CPR.

  \section{Supporting data on a Second Device}
  \label{sec:4termjj_second_device}

  \begin{figure}[t]
    \centering
    \includegraphics{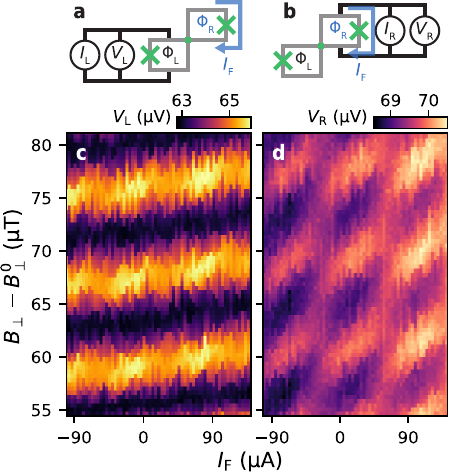}
    \caption{
      Fixed-Current SQUID Oscillations in Device B.
      \textbf{(a),(b)} SQUID oscillation measurement circuit for \textbf{(c)} and \textbf{(d)}, respectively.
      \textbf{(c),(d)} Fixed-current CPR Measurements in the resistive state of SQUID L \textbf{(c)} with $I_\mathrm{L}=\SI{1.45}{\micro\ampere}$ and SQUID R \textbf{(d)} with $I_\mathrm{R}=\SI{1.45}{\micro\ampere}$, with the opposite SQUID's leads floating.
      The magnetic field was swept near an approximate zero-field point $B_{\perp}^0=\SI{7.15}{\milli\tesla}$ determined from Fraunhofer pattern measurements of the reference Josephson junctions.
    }
    \label{fig:4termjj_device_B}
  \end{figure}

  Here we present a second set of nonlocal SQUID oscillation measurements akin to those in Fig.~\subref{fig:4termjj_anomalous}{(b)} and Fig.~\subref{fig:4termjj_squid_r_nonlocal}{(a)} on another device (Device B), with results shown in Fig.~\ref{fig:4termjj_device_B}.
  This device has an identical design to Device A and was fabricated on the same chip.
  Measurements for both SQUID L and SQUID R are presented in Figs.~\subref{fig:4termjj_device_B}{(c)} and \subref{fig:4termjj_device_B}{(d)}, respectively.
  The corresponding measurement circuits are shown in Figs.~\subref{fig:4termjj_device_B}{(a,b)}.
  In these measurements, with the opposite SQUID's leads floating, we observe $\Phi_0$-periodic oscillations in $V_\mathrm{L}$ primarily as a function of $B_{\perp}$ and in $V_\mathrm{R}$ both as a function of $B_{\perp}$ and $I_\mathrm{F}$, consistent with the flux-bias line almost exclusively affecting $\Phi_\mathrm{R}$ by design.
  In addition to the small direct cross coupling of $I_\mathrm{F}$ to $\Phi_\mathrm{L}$ giving the SQUID L oscillations a slight tilt in Fig.~\subref{fig:4termjj_device_B}{(c)}, periodic oscillations in the peak height are observed as a function of $I_\mathrm{F}$.
  The oscillations also exhibit a slight zig-zag pattern comparable to those of Device A seen in Fig.~\subref{fig:4termjj_anomalous}, but due to the significant jitter visible in the measurements, extracting a $\phi_0$ shift of the junction is difficult.
  The measurements of SQUID R are also qualitatively similar to those of Device A.

  \section{Two-terminal Junction Array Models of Multiterminal Junctions}
  \label{sec:4termjj_2tjj_models}

  Herein we derive expressions for the critical current of the four- and three-terminal JJ circuits shown in Fig.~\subref{fig:4termjj_2tjj_array}{(a-c)} by maximizing the supercurrent carried through the terminal labeled with an input current $I$.
  In both cases, we assume that the flux threading the multi-terminal JJ is negligible.
  We also assume that the reference junction critical currents are much larger than the critical currents of all two-terminal JJs describing the multiterminal JJ.
  For simplicity we neglect capacitances and linear inductances, and assume all junctions have CPRs of the form $I_\mathrm{c}\sin{(\phi)}$ where $\phi$ is the phase difference across the junction.

  We begin with the Andreev molecule device of Fig.~\subref{fig:4termjj_2tjj_array}{(b)}, which is designed to contain two JJs sharing a common superconducting terminal and separated by a distance on the order of the superconducting coherence length $\xi$, see Fig.~\subref{fig:4termjj_2tjj_array}{(d)}.
  To separately control the phase differences $\phi_\mathrm{A/B}$ across each junction, the two junctions are embedded in loops threaded by fluxes $\Phi_\mathrm{1}$ and $\Phi_\mathrm{2}$ (in units of $2\pi/\Phi_0$).
  Lastly, in one of the loops a reference junction of large critical current $I_\mathrm{c,1}^\mathrm{ref}$ is embedded to measure one of the JJ's CPRs.
  As these devices in previous experiments have been formed by connecting three superconducting contacts to a continuous region of conducting semiconducting material smaller than the coherence length \cite{Haxell_2023,Matsuo_2023_ArXiv_nonlocal}, it is feasible that supercurrent can directly travel between the two outermost terminals even in the absence of a central terminal.
  Consequently, we model the nonlocal coupling between the two JJs as a third JJ connecting the two outer terminals while bypassing the central one (gray in Fig.~\subref{fig:4termjj_2tjj_array}{(d)}.
  Notably, this model does not include wave function overlap between ABSs from different individual junctions.

  Given the phase differences across the junctions as defined in Fig.~\subref{fig:4termjj_2tjj_array}{(c)}, by flux quantization we have that $\phi_\mathrm{A}=\Phi_\mathrm{2}$.
  Next, since the critical current of the reference junction is very large compared to all others, its phase difference will adjust to whichever value maximizes the supercurrent through it, in this case $\pi/2$.
  Finally, by flux quantization we then have that $\pi/2-\phi_\mathrm{NL}=\Phi_\mathrm{1}$ and $\phi_\mathrm{NL}-\phi_\mathrm{B}-\phi_\mathrm{A}=0$.
  The signs of the phase differences are determined by an arbitrary but consistent definition of the current direction through each circuit branch \cite{Rasmussen_2021}, with fluxes defined as being associated with a clockwise current through a given loop.
  Accordingly, the critical current of the circuit is
  \begin{equation}\label{eq:haxell_molecule_ic}
    I_\mathrm{c} = I_\mathrm{c,1}^\mathrm{ref} + I_\mathrm{c,B}\cos{(\Phi_\mathrm{1}+\Phi_\mathrm{2})} + I_\mathrm{c,NL}\cos{(\Phi_\mathrm{1})}.
  \end{equation}

  Notably, the decision to approximate the flux $\Phi_1$ producing a phase difference across the shunting junction as opposed to directly tuning $\phi_\mathrm{A/B}$ was somewhat arbitrary.
  Modeling the circuit as the shunting junction existing out-of-plane such that $\Phi_1$ tunes $\phi_\mathrm{A/B}$ produces the same result as in eq.~\ref{eq:haxell_molecule_ic} except shifted by $\pi$ along the $\Phi_1$ axis.

  Proceeding to the circuit of Fig.~\subref{fig:4termjj_2tjj_array}{(d)} representing the device geometry of Ref.~\cite{Matsuo_2023_ArXiv_nonlocal}, we conduct similar calculations using flux quantization rules.
  In this case, there are reference junctions of critical currents $I_\mathrm{c,1}^\mathrm{ref}$ and $I_\mathrm{c,2}^\mathrm{ref}$ in the $\Phi_1$- and $\Phi_2$-threaded loops, respectively.
  With current $I$ being passed into the top left port and the others grounded, maximizing the supercurrent requires maximizing the current through reference junction 1, since we assume the reference junctions have arbitrarily large critical current.
  Hence, its phase will tend to $\pi/2$.
  For the other reference junction, it is connected to grounded terminals on both sides and so is not a bottleneck for the device's critical current.
  We can thus take the limit of infinite critical current such that no phase drop occurs across this junction, and we have $\phi_\mathrm{B} = \Phi_2$.
  Other flux quantization loops yield $\Phi_\mathrm{A}=\pi/2-\Phi_1$ and $\phi_\mathrm{NL}=\phi_\mathrm{A}-\phi_\mathrm{B}=\pi/2-\Phi_1-\Phi_2$.
  Hence, by calculating the current through each circuit branch connecting to the $I$ input, we find the device critical current to be
  \begin{equation}\label{eq:matsuo_molecule_ic}
    I_\mathrm{c} = I_\mathrm{c,1}^\mathrm{ref} + I_\mathrm{c,A}\cos{(\Phi_1)} + I_\mathrm{c,NL}\cos{(\Phi_1+\Phi_2)}.
  \end{equation}

  Lastly, we consider the 4TJJ circuit of Fig.~\subref{fig:4termjj_2tjj_array}{(a)}.
  Again, there are reference junctions of large critical currents $I_\mathrm{c,L}^\mathrm{ref}$ and $I_\mathrm{c,R}^\mathrm{ref}$ in the loops threaded by $\Phi_\mathrm{L}$ and $\Phi_\mathrm{R}$.
  The current $I$ flows through the branches containing reference junction L, the junction between terminals 1 and 2, and the junction between terminals 1 and 3.
  Since $I_\mathrm{c,L}^\mathrm{ref}$ is assumed very large, its phase at the critical current by the same reasoning as before is roughly $\pi/2$.
  Again, as in the previous case we can take $I_\mathrm{c,R}^\mathrm{ref}\rightarrow\infty$ since it is not a bottleneck for the critical current.
  From flux quantization, we then obtain $\phi_{12}=\Phi_\mathrm{L}+\pi/2$, $\phi_{34}=-\Phi_\mathrm{R}$, and $\phi_{13}+\phi_{34}+\phi_{42}-\phi_{12}=0$.
  To obtain enough equations to solve for all phases, we note that by Kirchhoff's current law we have $I_\mathrm{c,13}\sin{(\phi_{13})}=I_\mathrm{c,42}\sin{(\phi_{42})}$.
  Solving the last flux quantization condition for $\phi_{42}$ and substituting the result into the current conservation equation, we obtain
  \begin{equation}\label{eq:phi13_root_equation}
    I_\mathrm{c,13}\sin{(\phi_{13})} = I_\mathrm{c,42}\cos{(\Phi_\mathrm{L}+\Phi_\mathrm{R}-\phi_{13})}.
  \end{equation}
  This is a transcendental equation and has multiple solutions between $\phi_{13}\in[0,2\pi)$.
  The critical current of the circuit is by definition the maximum possible supercurrent it can sustain, so for every $(\Phi_\mathrm{L},\Phi_\mathrm{R})$ value we choose the $(\Phi_\mathrm{L},\Phi_\mathrm{R})$-dependent solution $\phi_{13}^\mathrm{max}$ to eq.~\ref{eq:phi13_root_equation} which maximizes $I$.
  The critical current of the circuit is then the sum of the current through the three paths branching from the input current $I$, given as
  \begin{equation}
    I_\mathrm{c} = I_\mathrm{c,L}^\mathrm{ref} + I_\mathrm{c,12}\cos{(\Phi_\mathrm{L})} + I_\mathrm{c,13}\sin{(\phi_{13}^\mathrm{max})}.
  \end{equation}
  For the calculated results in Fig.~\subref{fig:4termjj_2tjj_array}{(e)}, we plot them as a function of the flux generated by $I_\mathrm{F}$ and $B_\perp$, considering cross coupling of $I_\mathrm{F}$ into SQUID L determined from the oscillation periodicities of Fig.~\ref{fig:4termjj_characterization}.
  The $I_\mathrm{F}$ axis is converted into units of $\Phi_0$ by defining $1\times\Phi_0$ as a single flux quantum threading SQUID R due to $I_\mathrm{F}$ as well as the resulting cross coupling to SQUID L.
  For $B_\perp$, $1\times\Phi_0$ is defined as a single flux quantum threading both SQUIDs.

\end{appendix}

\bibliography{4termjj_bib.bib}

\begin{thebibliography}{58}%
\makeatletter
\providecommand \@ifxundefined [1]{%
 \@ifx{#1\undefined}
}%
\providecommand \@ifnum [1]{%
 \ifnum #1\expandafter \@firstoftwo
 \else \expandafter \@secondoftwo
 \fi
}%
\providecommand \@ifx [1]{%
 \ifx #1\expandafter \@firstoftwo
 \else \expandafter \@secondoftwo
 \fi
}%
\providecommand \natexlab [1]{#1}%
\providecommand \enquote  [1]{``#1''}%
\providecommand \bibnamefont  [1]{#1}%
\providecommand \bibfnamefont [1]{#1}%
\providecommand \citenamefont [1]{#1}%
\providecommand \href@noop [0]{\@secondoftwo}%
\providecommand \href [0]{\begingroup \@sanitize@url \@href}%
\providecommand \@href[1]{\@@startlink{#1}\@@href}%
\providecommand \@@href[1]{\endgroup#1\@@endlink}%
\providecommand \@sanitize@url [0]{\catcode `\\12\catcode `\$12\catcode
  `\&12\catcode `\#12\catcode `\^12\catcode `\_12\catcode `\%12\relax}%
\providecommand \@@startlink[1]{}%
\providecommand \@@endlink[0]{}%
\providecommand \url  [0]{\begingroup\@sanitize@url \@url }%
\providecommand \@url [1]{\endgroup\@href {#1}{\urlprefix }}%
\providecommand \urlprefix  [0]{URL }%
\providecommand \Eprint [0]{\href }%
\providecommand \doibase [0]{https://doi.org/}%
\providecommand \selectlanguage [0]{\@gobble}%
\providecommand \bibinfo  [0]{\@secondoftwo}%
\providecommand \bibfield  [0]{\@secondoftwo}%
\providecommand \translation [1]{[#1]}%
\providecommand \BibitemOpen [0]{}%
\providecommand \bibitemStop [0]{}%
\providecommand \bibitemNoStop [0]{.\EOS\space}%
\providecommand \EOS [0]{\spacefactor3000\relax}%
\providecommand \BibitemShut  [1]{\csname bibitem#1\endcsname}%
\let\auto@bib@innerbib\@empty
\bibitem [{\citenamefont {Beenakker}(1991)}]{Beenakker_1991}%
  \BibitemOpen
  \bibfield  {author} {\bibinfo {author} {\bibfnamefont {C.~W.~J.}\
  \bibnamefont {Beenakker}},\ }\bibfield  {title} {\bibinfo {title} {{Universal
  limit of critical-current fluctuations in mesoscopic Josephson junctions}},\
  }\href {https://doi.org/10.1103/physrevlett.67.3836} {\bibfield  {journal}
  {\bibinfo  {journal} {Physical Review Letters}\ }\textbf {\bibinfo {volume}
  {67}},\ \bibinfo {pages} {3836} (\bibinfo {year} {1991})}\BibitemShut
  {NoStop}%
\bibitem [{\citenamefont {van Heck}\ \emph {et~al.}(2014)\citenamefont {van
  Heck}, \citenamefont {Mi},\ and\ \citenamefont {Akhmerov}}]{Van_Heck_2014}%
  \BibitemOpen
  \bibfield  {author} {\bibinfo {author} {\bibfnamefont {B.}~\bibnamefont {van
  Heck}}, \bibinfo {author} {\bibfnamefont {S.}~\bibnamefont {Mi}},\ and\
  \bibinfo {author} {\bibfnamefont {A.~R.}\ \bibnamefont {Akhmerov}},\
  }\bibfield  {title} {\bibinfo {title} {{Single fermion manipulation via
  superconducting phase differences in multiterminal Josephson junctions}},\
  }\bibfield  {journal} {\bibinfo  {journal} {Physical Review B}\ }\textbf
  {\bibinfo {volume} {90}},\ \href {https://doi.org/10.1103/physrevb.90.155450}
  {10.1103/physrevb.90.155450} (\bibinfo {year} {2014})\BibitemShut {NoStop}%
\bibitem [{\citenamefont {Riwar}\ \emph {et~al.}(2016)\citenamefont {Riwar},
  \citenamefont {Houzet}, \citenamefont {Meyer},\ and\ \citenamefont
  {Nazarov}}]{Riwar_2016}%
  \BibitemOpen
  \bibfield  {author} {\bibinfo {author} {\bibfnamefont {R.-P.}\ \bibnamefont
  {Riwar}}, \bibinfo {author} {\bibfnamefont {M.}~\bibnamefont {Houzet}},
  \bibinfo {author} {\bibfnamefont {J.~S.}\ \bibnamefont {Meyer}},\ and\
  \bibinfo {author} {\bibfnamefont {Y.~V.}\ \bibnamefont {Nazarov}},\
  }\bibfield  {title} {\bibinfo {title} {Multi-terminal josephson junctions as
  topological matter},\ }\bibfield  {journal} {\bibinfo  {journal} {Nature
  Communications}\ }\textbf {\bibinfo {volume} {7}},\ \href
  {https://doi.org/10.1038/ncomms11167} {10.1038/ncomms11167} (\bibinfo {year}
  {2016})\BibitemShut {NoStop}%
\bibitem [{\citenamefont {Xie}\ \emph {et~al.}(2017)\citenamefont {Xie},
  \citenamefont {Vavilov},\ and\ \citenamefont {Levchenko}}]{Xie_2017}%
  \BibitemOpen
  \bibfield  {author} {\bibinfo {author} {\bibfnamefont {H.-Y.}\ \bibnamefont
  {Xie}}, \bibinfo {author} {\bibfnamefont {M.~G.}\ \bibnamefont {Vavilov}},\
  and\ \bibinfo {author} {\bibfnamefont {A.}~\bibnamefont {Levchenko}},\
  }\bibfield  {title} {\bibinfo {title} {Topological andreev bands in
  three-terminal josephson junctions},\ }\bibfield  {journal} {\bibinfo
  {journal} {Physical Review B}\ }\textbf {\bibinfo {volume} {96}},\ \href
  {https://doi.org/10.1103/physrevb.96.161406} {10.1103/physrevb.96.161406}
  (\bibinfo {year} {2017})\BibitemShut {NoStop}%
\bibitem [{\citenamefont {Houzet}\ and\ \citenamefont
  {Meyer}(2019)}]{Houzet_2019}%
  \BibitemOpen
  \bibfield  {author} {\bibinfo {author} {\bibfnamefont {M.}~\bibnamefont
  {Houzet}}\ and\ \bibinfo {author} {\bibfnamefont {J.~S.}\ \bibnamefont
  {Meyer}},\ }\bibfield  {title} {\bibinfo {title} {{Majorana-Weyl crossings in
  topological multiterminal junctions}},\ }\bibfield  {journal} {\bibinfo
  {journal} {Physical Review B}\ }\textbf {\bibinfo {volume} {100}},\ \href
  {https://doi.org/10.1103/physrevb.100.014521} {10.1103/physrevb.100.014521}
  (\bibinfo {year} {2019})\BibitemShut {NoStop}%
\bibitem [{\citenamefont {Fatemi}\ \emph {et~al.}(2021)\citenamefont {Fatemi},
  \citenamefont {Akhmerov},\ and\ \citenamefont {Bretheau}}]{Fatemi_2021}%
  \BibitemOpen
  \bibfield  {author} {\bibinfo {author} {\bibfnamefont {V.}~\bibnamefont
  {Fatemi}}, \bibinfo {author} {\bibfnamefont {A.~R.}\ \bibnamefont
  {Akhmerov}},\ and\ \bibinfo {author} {\bibfnamefont {L.}~\bibnamefont
  {Bretheau}},\ }\bibfield  {title} {\bibinfo {title} {{Weyl Josephson
  circuits}},\ }\bibfield  {journal} {\bibinfo  {journal} {Physical Review
  Research}\ }\textbf {\bibinfo {volume} {3}},\ \href
  {https://doi.org/10.1103/physrevresearch.3.013288}
  {10.1103/physrevresearch.3.013288} (\bibinfo {year} {2021})\BibitemShut
  {NoStop}%
\bibitem [{\citenamefont {Barakov}\ and\ \citenamefont
  {Nazarov}(2023)}]{Barakov_2023}%
  \BibitemOpen
  \bibfield  {author} {\bibinfo {author} {\bibfnamefont {H.}~\bibnamefont
  {Barakov}}\ and\ \bibinfo {author} {\bibfnamefont {Y.~V.}\ \bibnamefont
  {Nazarov}},\ }\bibfield  {title} {\bibinfo {title} {{Abundance of Weyl points
  in semiclassical multiterminal superconducting nanostructures}},\ }\bibfield
  {journal} {\bibinfo  {journal} {Physical Review B}\ }\textbf {\bibinfo
  {volume} {107}},\ \href {https://doi.org/10.1103/physrevb.107.014507}
  {10.1103/physrevb.107.014507} (\bibinfo {year} {2023})\BibitemShut {NoStop}%
\bibitem [{\citenamefont {Meyer}\ and\ \citenamefont
  {Houzet}(2017)}]{Meyer_2017}%
  \BibitemOpen
  \bibfield  {author} {\bibinfo {author} {\bibfnamefont {J.~S.}\ \bibnamefont
  {Meyer}}\ and\ \bibinfo {author} {\bibfnamefont {M.}~\bibnamefont {Houzet}},\
  }\bibfield  {title} {\bibinfo {title} {{Nontrivial Chern Numbers in
  Three-Terminal Josephson Junctions}},\ }\bibfield  {journal} {\bibinfo
  {journal} {Physical Review Letters}\ }\textbf {\bibinfo {volume} {119}},\
  \href {https://doi.org/10.1103/physrevlett.119.136807}
  {10.1103/physrevlett.119.136807} (\bibinfo {year} {2017})\BibitemShut
  {NoStop}%
\bibitem [{\citenamefont {de~Bruyn~Ouboter}\ \emph {et~al.}(1995)\citenamefont
  {de~Bruyn~Ouboter}, \citenamefont {Omelyanchouk},\ and\ \citenamefont
  {Vol}}]{de_Bruyn_Ouboter_1995}%
  \BibitemOpen
  \bibfield  {author} {\bibinfo {author} {\bibfnamefont {R.}~\bibnamefont
  {de~Bruyn~Ouboter}}, \bibinfo {author} {\bibfnamefont {A.}~\bibnamefont
  {Omelyanchouk}},\ and\ \bibinfo {author} {\bibfnamefont {E.}~\bibnamefont
  {Vol}},\ }\bibfield  {title} {\bibinfo {title} {{Multi-terminal {SQUID}
  controlled by the transport current}},\ }\href
  {https://doi.org/10.1016/0921-4526(94)00299-b} {\bibfield  {journal}
  {\bibinfo  {journal} {Physica B: Condensed Matter}\ }\textbf {\bibinfo
  {volume} {205}},\ \bibinfo {pages} {153} (\bibinfo {year}
  {1995})}\BibitemShut {NoStop}%
\bibitem [{\citenamefont {Zareyan}\ and\ \citenamefont
  {Omelyanchuk}(1999)}]{Zareyan_1999}%
  \BibitemOpen
  \bibfield  {author} {\bibinfo {author} {\bibfnamefont {M.}~\bibnamefont
  {Zareyan}}\ and\ \bibinfo {author} {\bibfnamefont {A.~N.}\ \bibnamefont
  {Omelyanchuk}},\ }\bibfield  {title} {\bibinfo {title} {{Coherent current
  states in a mesoscopic four-terminal Josephson junction}},\ }\href
  {https://doi.org/10.1063/1.593723} {\bibfield  {journal} {\bibinfo  {journal}
  {Low Temperature Physics}\ }\textbf {\bibinfo {volume} {25}},\ \bibinfo
  {pages} {175} (\bibinfo {year} {1999})}\BibitemShut {NoStop}%
\bibitem [{\citenamefont {Omelyanchouk}\ and\ \citenamefont
  {Zareyan}(2000)}]{Omelyanchouk_2000}%
  \BibitemOpen
  \bibfield  {author} {\bibinfo {author} {\bibfnamefont {A.}~\bibnamefont
  {Omelyanchouk}}\ and\ \bibinfo {author} {\bibfnamefont {M.}~\bibnamefont
  {Zareyan}},\ }\bibfield  {title} {\bibinfo {title} {{Ballistic four-terminal
  Josephson junction: bistable states and magnetic flux transfer}},\ }\href
  {https://doi.org/10.1016/s0921-4526(99)00760-7} {\bibfield  {journal}
  {\bibinfo  {journal} {Physica B: Condensed Matter}\ }\textbf {\bibinfo
  {volume} {291}},\ \bibinfo {pages} {81} (\bibinfo {year} {2000})}\BibitemShut
  {NoStop}%
\bibitem [{\citenamefont {Amin}\ \emph {et~al.}(2001)\citenamefont {Amin},
  \citenamefont {Omelyanchouk},\ and\ \citenamefont {Zagoskin}}]{Amin_2001}%
  \BibitemOpen
  \bibfield  {author} {\bibinfo {author} {\bibfnamefont {M.~H.~S.}\
  \bibnamefont {Amin}}, \bibinfo {author} {\bibfnamefont {A.~N.}\ \bibnamefont
  {Omelyanchouk}},\ and\ \bibinfo {author} {\bibfnamefont {A.~M.}\ \bibnamefont
  {Zagoskin}},\ }\bibfield  {title} {\bibinfo {title} {{Mesoscopic
  multiterminal Josephson structures. I. Effects of nonlocal weak coupling}},\
  }\href {https://doi.org/10.1063/1.1399198} {\bibfield  {journal} {\bibinfo
  {journal} {Low Temperature Physics}\ }\textbf {\bibinfo {volume} {27}},\
  \bibinfo {pages} {616} (\bibinfo {year} {2001})}\BibitemShut {NoStop}%
\bibitem [{\citenamefont {Alidoust}\ \emph {et~al.}(2012)\citenamefont
  {Alidoust}, \citenamefont {Sewell},\ and\ \citenamefont
  {Linder}}]{Alidoust_2012}%
  \BibitemOpen
  \bibfield  {author} {\bibinfo {author} {\bibfnamefont {M.}~\bibnamefont
  {Alidoust}}, \bibinfo {author} {\bibfnamefont {G.}~\bibnamefont {Sewell}},\
  and\ \bibinfo {author} {\bibfnamefont {J.}~\bibnamefont {Linder}},\
  }\bibfield  {title} {\bibinfo {title} {{Superconducting phase transistor in
  diffusive four-terminal ferromagnetic Josephson junctions}},\ }\bibfield
  {journal} {\bibinfo  {journal} {Physical Review B}\ }\textbf {\bibinfo
  {volume} {85}},\ \href {https://doi.org/10.1103/physrevb.85.144520}
  {10.1103/physrevb.85.144520} (\bibinfo {year} {2012})\BibitemShut {NoStop}%
\bibitem [{\citenamefont {Amin}\ \emph {et~al.}(2002)\citenamefont {Amin},
  \citenamefont {Omelyanchouk}, \citenamefont {Blais}, \citenamefont {van~den
  Brink}, \citenamefont {Rose}, \citenamefont {Duty},\ and\ \citenamefont
  {Zagoskin}}]{Amin_2002}%
  \BibitemOpen
  \bibfield  {author} {\bibinfo {author} {\bibfnamefont {M.}~\bibnamefont
  {Amin}}, \bibinfo {author} {\bibfnamefont {A.}~\bibnamefont {Omelyanchouk}},
  \bibinfo {author} {\bibfnamefont {A.}~\bibnamefont {Blais}}, \bibinfo
  {author} {\bibfnamefont {A.~M.}\ \bibnamefont {van~den Brink}}, \bibinfo
  {author} {\bibfnamefont {G.}~\bibnamefont {Rose}}, \bibinfo {author}
  {\bibfnamefont {T.}~\bibnamefont {Duty}},\ and\ \bibinfo {author}
  {\bibfnamefont {A.}~\bibnamefont {Zagoskin}},\ }\bibfield  {title} {\bibinfo
  {title} {{Multi-terminal superconducting phase qubit}},\ }\href
  {https://doi.org/10.1016/s0921-4534(01)01187-x} {\bibfield  {journal}
  {\bibinfo  {journal} {Physica C: Superconductivity}\ }\textbf {\bibinfo
  {volume} {368}},\ \bibinfo {pages} {310} (\bibinfo {year}
  {2002})}\BibitemShut {NoStop}%
\bibitem [{\citenamefont {Vleeming}\ \emph {et~al.}(1996)\citenamefont
  {Vleeming}, \citenamefont {Zakarian}, \citenamefont {Omelyanchouk},\ and\
  \citenamefont {de~Bruyn~Ouboter}}]{Vleeming_1996}%
  \BibitemOpen
  \bibfield  {author} {\bibinfo {author} {\bibfnamefont {B.}~\bibnamefont
  {Vleeming}}, \bibinfo {author} {\bibfnamefont {A.}~\bibnamefont {Zakarian}},
  \bibinfo {author} {\bibfnamefont {A.}~\bibnamefont {Omelyanchouk}},\ and\
  \bibinfo {author} {\bibfnamefont {R.}~\bibnamefont {de~Bruyn~Ouboter}},\
  }\bibfield  {title} {\bibinfo {title} {Macroscopic quantum interference
  phenomena in a 4-terminal {SQUID}},\ }\href
  {https://doi.org/10.1016/0921-4526(96)00485-1} {\bibfield  {journal}
  {\bibinfo  {journal} {Physica B: Condensed Matter}\ }\textbf {\bibinfo
  {volume} {226}},\ \bibinfo {pages} {253} (\bibinfo {year}
  {1996})}\BibitemShut {NoStop}%
\bibitem [{\citenamefont {Vleeming}\ \emph {et~al.}(1999)\citenamefont
  {Vleeming}, \citenamefont {van Bemmelen}, \citenamefont {Berends},
  \citenamefont {Ouboter},\ and\ \citenamefont {Omelyanchouk}}]{Vleeming_1999}%
  \BibitemOpen
  \bibfield  {author} {\bibinfo {author} {\bibfnamefont {B.}~\bibnamefont
  {Vleeming}}, \bibinfo {author} {\bibfnamefont {F.}~\bibnamefont {van
  Bemmelen}}, \bibinfo {author} {\bibfnamefont {M.}~\bibnamefont {Berends}},
  \bibinfo {author} {\bibfnamefont {R.~B.}\ \bibnamefont {Ouboter}},\ and\
  \bibinfo {author} {\bibfnamefont {A.}~\bibnamefont {Omelyanchouk}},\
  }\bibfield  {title} {\bibinfo {title} {{Measurements of the flux, embraced by
  the ring of a four-terminal {SQUID}, as a function of the external magnetic
  flux and the applied transport current}},\ }\href
  {https://doi.org/10.1016/s0921-4526(98)01122-3} {\bibfield  {journal}
  {\bibinfo  {journal} {Physica B: Condensed Matter}\ }\textbf {\bibinfo
  {volume} {262}},\ \bibinfo {pages} {296} (\bibinfo {year}
  {1999})}\BibitemShut {NoStop}%
\bibitem [{\citenamefont {Draelos}\ \emph {et~al.}(2019)\citenamefont
  {Draelos}, \citenamefont {Wei}, \citenamefont {Seredinski}, \citenamefont
  {Li}, \citenamefont {Mehta}, \citenamefont {Watanabe}, \citenamefont
  {Taniguchi}, \citenamefont {Borzenets}, \citenamefont {Amet},\ and\
  \citenamefont {Finkelstein}}]{Draelos_2019}%
  \BibitemOpen
  \bibfield  {author} {\bibinfo {author} {\bibfnamefont {A.~W.}\ \bibnamefont
  {Draelos}}, \bibinfo {author} {\bibfnamefont {M.-T.}\ \bibnamefont {Wei}},
  \bibinfo {author} {\bibfnamefont {A.}~\bibnamefont {Seredinski}}, \bibinfo
  {author} {\bibfnamefont {H.}~\bibnamefont {Li}}, \bibinfo {author}
  {\bibfnamefont {Y.}~\bibnamefont {Mehta}}, \bibinfo {author} {\bibfnamefont
  {K.}~\bibnamefont {Watanabe}}, \bibinfo {author} {\bibfnamefont
  {T.}~\bibnamefont {Taniguchi}}, \bibinfo {author} {\bibfnamefont {I.~V.}\
  \bibnamefont {Borzenets}}, \bibinfo {author} {\bibfnamefont {F.}~\bibnamefont
  {Amet}},\ and\ \bibinfo {author} {\bibfnamefont {G.}~\bibnamefont
  {Finkelstein}},\ }\bibfield  {title} {\bibinfo {title} {{Supercurrent Flow in
  Multiterminal Graphene Josephson Junctions}},\ }\href
  {https://doi.org/10.1021/acs.nanolett.8b04330} {\bibfield  {journal}
  {\bibinfo  {journal} {Nano Letters}\ }\textbf {\bibinfo {volume} {19}},\
  \bibinfo {pages} {1039} (\bibinfo {year} {2019})}\BibitemShut {NoStop}%
\bibitem [{\citenamefont {Pankratova}\ \emph {et~al.}(2020)\citenamefont
  {Pankratova}, \citenamefont {Lee}, \citenamefont {Kuzmin}, \citenamefont
  {Wickramasinghe}, \citenamefont {Mayer}, \citenamefont {Yuan}, \citenamefont
  {Vavilov}, \citenamefont {Shabani},\ and\ \citenamefont
  {Manucharyan}}]{Pankratova_2020}%
  \BibitemOpen
  \bibfield  {author} {\bibinfo {author} {\bibfnamefont {N.}~\bibnamefont
  {Pankratova}}, \bibinfo {author} {\bibfnamefont {H.}~\bibnamefont {Lee}},
  \bibinfo {author} {\bibfnamefont {R.}~\bibnamefont {Kuzmin}}, \bibinfo
  {author} {\bibfnamefont {K.}~\bibnamefont {Wickramasinghe}}, \bibinfo
  {author} {\bibfnamefont {W.}~\bibnamefont {Mayer}}, \bibinfo {author}
  {\bibfnamefont {J.}~\bibnamefont {Yuan}}, \bibinfo {author} {\bibfnamefont
  {M.~G.}\ \bibnamefont {Vavilov}}, \bibinfo {author} {\bibfnamefont
  {J.}~\bibnamefont {Shabani}},\ and\ \bibinfo {author} {\bibfnamefont {V.~E.}\
  \bibnamefont {Manucharyan}},\ }\bibfield  {title} {\bibinfo {title}
  {Multiterminal josephson effect},\ }\bibfield  {journal} {\bibinfo  {journal}
  {Physical Review X}\ }\textbf {\bibinfo {volume} {10}},\ \href
  {https://doi.org/10.1103/physrevx.10.031051} {10.1103/physrevx.10.031051}
  (\bibinfo {year} {2020})\BibitemShut {NoStop}%
\bibitem [{\citenamefont {Graziano}\ \emph {et~al.}(2020)\citenamefont
  {Graziano}, \citenamefont {Lee}, \citenamefont {Pendharkar}, \citenamefont
  {Palmstr{\o}m},\ and\ \citenamefont {Pribiag}}]{Graziano_2020}%
  \BibitemOpen
  \bibfield  {author} {\bibinfo {author} {\bibfnamefont {G.~V.}\ \bibnamefont
  {Graziano}}, \bibinfo {author} {\bibfnamefont {J.~S.}\ \bibnamefont {Lee}},
  \bibinfo {author} {\bibfnamefont {M.}~\bibnamefont {Pendharkar}}, \bibinfo
  {author} {\bibfnamefont {C.~J.}\ \bibnamefont {Palmstr{\o}m}},\ and\ \bibinfo
  {author} {\bibfnamefont {V.~S.}\ \bibnamefont {Pribiag}},\ }\bibfield
  {title} {\bibinfo {title} {{Transport studies in a gate-tunable
  three-terminal Josephson junction}},\ }\bibfield  {journal} {\bibinfo
  {journal} {Physical Review B}\ }\textbf {\bibinfo {volume} {101}},\ \href
  {https://doi.org/10.1103/physrevb.101.054510} {10.1103/physrevb.101.054510}
  (\bibinfo {year} {2020})\BibitemShut {NoStop}%
\bibitem [{\citenamefont {Arnault}\ \emph {et~al.}(2021)\citenamefont
  {Arnault}, \citenamefont {Larson}, \citenamefont {Seredinski}, \citenamefont
  {Zhao}, \citenamefont {Idris}, \citenamefont {McConnell}, \citenamefont
  {Watanabe}, \citenamefont {Taniguchi}, \citenamefont {Borzenets},
  \citenamefont {Amet},\ and\ \citenamefont {Finkelstein}}]{Arnault_2021}%
  \BibitemOpen
  \bibfield  {author} {\bibinfo {author} {\bibfnamefont {E.~G.}\ \bibnamefont
  {Arnault}}, \bibinfo {author} {\bibfnamefont {T.~F.~Q.}\ \bibnamefont
  {Larson}}, \bibinfo {author} {\bibfnamefont {A.}~\bibnamefont {Seredinski}},
  \bibinfo {author} {\bibfnamefont {L.}~\bibnamefont {Zhao}}, \bibinfo {author}
  {\bibfnamefont {S.}~\bibnamefont {Idris}}, \bibinfo {author} {\bibfnamefont
  {A.}~\bibnamefont {McConnell}}, \bibinfo {author} {\bibfnamefont
  {K.}~\bibnamefont {Watanabe}}, \bibinfo {author} {\bibfnamefont
  {T.}~\bibnamefont {Taniguchi}}, \bibinfo {author} {\bibfnamefont
  {I.}~\bibnamefont {Borzenets}}, \bibinfo {author} {\bibfnamefont
  {F.}~\bibnamefont {Amet}},\ and\ \bibinfo {author} {\bibfnamefont
  {G.}~\bibnamefont {Finkelstein}},\ }\bibfield  {title} {\bibinfo {title}
  {Multiterminal inverse {AC} josephson effect},\ }\href
  {https://doi.org/10.1021/acs.nanolett.1c03474} {\bibfield  {journal}
  {\bibinfo  {journal} {Nano Letters}\ }\textbf {\bibinfo {volume} {21}},\
  \bibinfo {pages} {9668} (\bibinfo {year} {2021})}\BibitemShut {NoStop}%
\bibitem [{\citenamefont {Graziano}\ \emph {et~al.}(2022)\citenamefont
  {Graziano}, \citenamefont {Gupta}, \citenamefont {Pendharkar}, \citenamefont
  {Dong}, \citenamefont {Dempsey}, \citenamefont {Palmstr{\o}m},\ and\
  \citenamefont {Pribiag}}]{Graziano_2022}%
  \BibitemOpen
  \bibfield  {author} {\bibinfo {author} {\bibfnamefont {G.~V.}\ \bibnamefont
  {Graziano}}, \bibinfo {author} {\bibfnamefont {M.}~\bibnamefont {Gupta}},
  \bibinfo {author} {\bibfnamefont {M.}~\bibnamefont {Pendharkar}}, \bibinfo
  {author} {\bibfnamefont {J.~T.}\ \bibnamefont {Dong}}, \bibinfo {author}
  {\bibfnamefont {C.~P.}\ \bibnamefont {Dempsey}}, \bibinfo {author}
  {\bibfnamefont {C.}~\bibnamefont {Palmstr{\o}m}},\ and\ \bibinfo {author}
  {\bibfnamefont {V.~S.}\ \bibnamefont {Pribiag}},\ }\bibfield  {title}
  {\bibinfo {title} {Selective control of conductance modes in multi-terminal
  josephson junctions},\ }\bibfield  {journal} {\bibinfo  {journal} {Nature
  Communications}\ }\textbf {\bibinfo {volume} {13}},\ \href
  {https://doi.org/10.1038/s41467-022-33682-2} {10.1038/s41467-022-33682-2}
  (\bibinfo {year} {2022})\BibitemShut {NoStop}%
\bibitem [{\citenamefont {Lee}(2022)}]{Lee_2022_thesis}%
  \BibitemOpen
  \bibfield  {author} {\bibinfo {author} {\bibfnamefont {H.}~\bibnamefont
  {Lee}},\ }\emph {\bibinfo {title} {{Supercurrent and Andreev bound states in
  multi-terminal Josephson junctions}}},\ \href@noop {} {Ph.D. thesis},\
  \bibinfo  {school} {University of Maryland} (\bibinfo {year}
  {2022})\BibitemShut {NoStop}%
\bibitem [{\citenamefont {Chiles}\ \emph {et~al.}(2023)\citenamefont {Chiles},
  \citenamefont {Arnault}, \citenamefont {Chen}, \citenamefont {Larson},
  \citenamefont {Zhao}, \citenamefont {Watanabe}, \citenamefont {Taniguchi},
  \citenamefont {Amet},\ and\ \citenamefont {Finkelstein}}]{Chiles_2023}%
  \BibitemOpen
  \bibfield  {author} {\bibinfo {author} {\bibfnamefont {J.}~\bibnamefont
  {Chiles}}, \bibinfo {author} {\bibfnamefont {E.~G.}\ \bibnamefont {Arnault}},
  \bibinfo {author} {\bibfnamefont {C.-C.}\ \bibnamefont {Chen}}, \bibinfo
  {author} {\bibfnamefont {T.~F.~Q.}\ \bibnamefont {Larson}}, \bibinfo {author}
  {\bibfnamefont {L.}~\bibnamefont {Zhao}}, \bibinfo {author} {\bibfnamefont
  {K.}~\bibnamefont {Watanabe}}, \bibinfo {author} {\bibfnamefont
  {T.}~\bibnamefont {Taniguchi}}, \bibinfo {author} {\bibfnamefont
  {F.}~\bibnamefont {Amet}},\ and\ \bibinfo {author} {\bibfnamefont
  {G.}~\bibnamefont {Finkelstein}},\ }\bibfield  {title} {\bibinfo {title}
  {{Nonreciprocal Supercurrents in a Field-Free Graphene Josephson Triode}},\
  }\href {https://doi.org/10.1021/acs.nanolett.3c01276} {\bibfield  {journal}
  {\bibinfo  {journal} {Nano Letters}\ }\textbf {\bibinfo {volume} {23}},\
  \bibinfo {pages} {5257} (\bibinfo {year} {2023})}\BibitemShut {NoStop}%
\bibitem [{\citenamefont {Pfeffer}\ \emph {et~al.}(2014)\citenamefont
  {Pfeffer}, \citenamefont {Duvauchelle}, \citenamefont {Courtois},
  \citenamefont {M{\'{e}}lin}, \citenamefont {Feinberg},\ and\ \citenamefont
  {Lefloch}}]{Pfeffer_2014}%
  \BibitemOpen
  \bibfield  {author} {\bibinfo {author} {\bibfnamefont {A.~H.}\ \bibnamefont
  {Pfeffer}}, \bibinfo {author} {\bibfnamefont {J.~E.}\ \bibnamefont
  {Duvauchelle}}, \bibinfo {author} {\bibfnamefont {H.}~\bibnamefont
  {Courtois}}, \bibinfo {author} {\bibfnamefont {R.}~\bibnamefont
  {M{\'{e}}lin}}, \bibinfo {author} {\bibfnamefont {D.}~\bibnamefont
  {Feinberg}},\ and\ \bibinfo {author} {\bibfnamefont {F.}~\bibnamefont
  {Lefloch}},\ }\bibfield  {title} {\bibinfo {title} {Subgap structure in the
  conductance of a three-terminal josephson junction},\ }\bibfield  {journal}
  {\bibinfo  {journal} {Physical Review B}\ }\textbf {\bibinfo {volume} {90}},\
  \href {https://doi.org/10.1103/physrevb.90.075401}
  {10.1103/physrevb.90.075401} (\bibinfo {year} {2014})\BibitemShut {NoStop}%
\bibitem [{\citenamefont {Cohen}\ \emph {et~al.}(2018)\citenamefont {Cohen},
  \citenamefont {Ronen}, \citenamefont {Kang}, \citenamefont {Heiblum},
  \citenamefont {Feinberg}, \citenamefont {M{\'{e}}lin},\ and\ \citenamefont
  {Shtrikman}}]{Cohen_2018}%
  \BibitemOpen
  \bibfield  {author} {\bibinfo {author} {\bibfnamefont {Y.}~\bibnamefont
  {Cohen}}, \bibinfo {author} {\bibfnamefont {Y.}~\bibnamefont {Ronen}},
  \bibinfo {author} {\bibfnamefont {J.-H.}\ \bibnamefont {Kang}}, \bibinfo
  {author} {\bibfnamefont {M.}~\bibnamefont {Heiblum}}, \bibinfo {author}
  {\bibfnamefont {D.}~\bibnamefont {Feinberg}}, \bibinfo {author}
  {\bibfnamefont {R.}~\bibnamefont {M{\'{e}}lin}},\ and\ \bibinfo {author}
  {\bibfnamefont {H.}~\bibnamefont {Shtrikman}},\ }\bibfield  {title} {\bibinfo
  {title} {{Nonlocal supercurrent of quartets in a three-terminal Josephson
  junction}},\ }\href {https://doi.org/10.1073/pnas.1800044115} {\bibfield
  {journal} {\bibinfo  {journal} {Proceedings of the National Academy of
  Sciences}\ }\textbf {\bibinfo {volume} {115}},\ \bibinfo {pages} {6991}
  (\bibinfo {year} {2018})}\BibitemShut {NoStop}%
\bibitem [{\citenamefont {Huang}\ \emph {et~al.}(2022)\citenamefont {Huang},
  \citenamefont {Ronen}, \citenamefont {M{\'{e}}lin}, \citenamefont {Feinberg},
  \citenamefont {Watanabe}, \citenamefont {Taniguchi},\ and\ \citenamefont
  {Kim}}]{Huang_2022}%
  \BibitemOpen
  \bibfield  {author} {\bibinfo {author} {\bibfnamefont {K.-F.}\ \bibnamefont
  {Huang}}, \bibinfo {author} {\bibfnamefont {Y.}~\bibnamefont {Ronen}},
  \bibinfo {author} {\bibfnamefont {R.}~\bibnamefont {M{\'{e}}lin}}, \bibinfo
  {author} {\bibfnamefont {D.}~\bibnamefont {Feinberg}}, \bibinfo {author}
  {\bibfnamefont {K.}~\bibnamefont {Watanabe}}, \bibinfo {author}
  {\bibfnamefont {T.}~\bibnamefont {Taniguchi}},\ and\ \bibinfo {author}
  {\bibfnamefont {P.}~\bibnamefont {Kim}},\ }\bibfield  {title} {\bibinfo
  {title} {{Evidence for 4e charge of Cooper quartets in a biased
  multi-terminal graphene-based Josephson junction}},\ }\bibfield  {journal}
  {\bibinfo  {journal} {Nature Communications}\ }\textbf {\bibinfo {volume}
  {13}},\ \href {https://doi.org/10.1038/s41467-022-30732-7}
  {10.1038/s41467-022-30732-7} (\bibinfo {year} {2022})\BibitemShut {NoStop}%
\bibitem [{\citenamefont {Zhang}\ \emph {et~al.}(2023)\citenamefont {Zhang},
  \citenamefont {Rashid}, \citenamefont {Ahari}, \citenamefont {Zhang},
  \citenamefont {Ananthanarayanan}, \citenamefont {Xiao}, \citenamefont
  {de~Coster}, \citenamefont {Gilbert}, \citenamefont {Samarth},\ and\
  \citenamefont {Kayyalha}}]{Zhang_2023}%
  \BibitemOpen
  \bibfield  {author} {\bibinfo {author} {\bibfnamefont {F.}~\bibnamefont
  {Zhang}}, \bibinfo {author} {\bibfnamefont {A.~S.}\ \bibnamefont {Rashid}},
  \bibinfo {author} {\bibfnamefont {M.~T.}\ \bibnamefont {Ahari}}, \bibinfo
  {author} {\bibfnamefont {W.}~\bibnamefont {Zhang}}, \bibinfo {author}
  {\bibfnamefont {K.~M.}\ \bibnamefont {Ananthanarayanan}}, \bibinfo {author}
  {\bibfnamefont {R.}~\bibnamefont {Xiao}}, \bibinfo {author} {\bibfnamefont
  {G.~J.}\ \bibnamefont {de~Coster}}, \bibinfo {author} {\bibfnamefont {M.~J.}\
  \bibnamefont {Gilbert}}, \bibinfo {author} {\bibfnamefont {N.}~\bibnamefont
  {Samarth}},\ and\ \bibinfo {author} {\bibfnamefont {M.}~\bibnamefont
  {Kayyalha}},\ }\bibfield  {title} {\bibinfo {title} {Andreev processes in
  mesoscopic multiterminal graphene josephson junctions},\ }\bibfield
  {journal} {\bibinfo  {journal} {Physical Review B}\ }\textbf {\bibinfo
  {volume} {107}},\ \href {https://doi.org/10.1103/physrevb.107.l140503}
  {10.1103/physrevb.107.l140503} (\bibinfo {year} {2023})\BibitemShut {NoStop}%
\bibitem [{\citenamefont {Gupta}\ \emph {et~al.}(2023)\citenamefont {Gupta},
  \citenamefont {Graziano}, \citenamefont {Pendharkar}, \citenamefont {Dong},
  \citenamefont {Dempsey}, \citenamefont {Palmstr{\o}m},\ and\ \citenamefont
  {Pribiag}}]{Gupta_2023}%
  \BibitemOpen
  \bibfield  {author} {\bibinfo {author} {\bibfnamefont {M.}~\bibnamefont
  {Gupta}}, \bibinfo {author} {\bibfnamefont {G.~V.}\ \bibnamefont {Graziano}},
  \bibinfo {author} {\bibfnamefont {M.}~\bibnamefont {Pendharkar}}, \bibinfo
  {author} {\bibfnamefont {J.~T.}\ \bibnamefont {Dong}}, \bibinfo {author}
  {\bibfnamefont {C.~P.}\ \bibnamefont {Dempsey}}, \bibinfo {author}
  {\bibfnamefont {C.}~\bibnamefont {Palmstr{\o}m}},\ and\ \bibinfo {author}
  {\bibfnamefont {V.~S.}\ \bibnamefont {Pribiag}},\ }\bibfield  {title}
  {\bibinfo {title} {{Gate-tunable superconducting diode effect in a
  three-terminal Josephson device}},\ }\bibfield  {journal} {\bibinfo
  {journal} {Nature Communications}\ }\textbf {\bibinfo {volume} {14}},\ \href
  {https://doi.org/10.1038/s41467-023-38856-0} {10.1038/s41467-023-38856-0}
  (\bibinfo {year} {2023})\BibitemShut {NoStop}%
\bibitem [{\citenamefont {Coraiola}\ \emph
  {et~al.}(2023{\natexlab{a}})\citenamefont {Coraiola}, \citenamefont {Haxell},
  \citenamefont {Sabonis}, \citenamefont {Weisbrich}, \citenamefont
  {Svetogorov}, \citenamefont {Hinderling}, \citenamefont {ten Kate},
  \citenamefont {Cheah}, \citenamefont {Krizek}, \citenamefont {Schott},
  \citenamefont {Wegscheider}, \citenamefont {Cuevas}, \citenamefont {Belzig},\
  and\ \citenamefont {Nichele}}]{Coraiola_2023}%
  \BibitemOpen
  \bibfield  {author} {\bibinfo {author} {\bibfnamefont {M.}~\bibnamefont
  {Coraiola}}, \bibinfo {author} {\bibfnamefont {D.~Z.}\ \bibnamefont
  {Haxell}}, \bibinfo {author} {\bibfnamefont {D.}~\bibnamefont {Sabonis}},
  \bibinfo {author} {\bibfnamefont {H.}~\bibnamefont {Weisbrich}}, \bibinfo
  {author} {\bibfnamefont {A.~E.}\ \bibnamefont {Svetogorov}}, \bibinfo
  {author} {\bibfnamefont {M.}~\bibnamefont {Hinderling}}, \bibinfo {author}
  {\bibfnamefont {S.~C.}\ \bibnamefont {ten Kate}}, \bibinfo {author}
  {\bibfnamefont {E.}~\bibnamefont {Cheah}}, \bibinfo {author} {\bibfnamefont
  {F.}~\bibnamefont {Krizek}}, \bibinfo {author} {\bibfnamefont
  {R.}~\bibnamefont {Schott}}, \bibinfo {author} {\bibfnamefont
  {W.}~\bibnamefont {Wegscheider}}, \bibinfo {author} {\bibfnamefont {J.~C.}\
  \bibnamefont {Cuevas}}, \bibinfo {author} {\bibfnamefont {W.}~\bibnamefont
  {Belzig}},\ and\ \bibinfo {author} {\bibfnamefont {F.}~\bibnamefont
  {Nichele}},\ }\bibfield  {title} {\bibinfo {title} {Phase-engineering the
  andreev band structure of a three-terminal josephson junction},\ }\bibfield
  {journal} {\bibinfo  {journal} {Nature Communications}\ }\textbf {\bibinfo
  {volume} {14}},\ \href {https://doi.org/10.1038/s41467-023-42356-6}
  {10.1038/s41467-023-42356-6} (\bibinfo {year}
  {2023}{\natexlab{a}})\BibitemShut {NoStop}%
\bibitem [{\citenamefont {{Matsuo}}\ \emph {et~al.}(2023)\citenamefont
  {{Matsuo}}, \citenamefont {{Imoto}}, \citenamefont {{Yokoyama}},
  \citenamefont {{Sato}}, \citenamefont {{Lindemann}}, \citenamefont
  {{Gronin}}, \citenamefont {{Gardner}}, \citenamefont {{Nakosai}},
  \citenamefont {{Tanaka}}, \citenamefont {{Manfra}},\ and\ \citenamefont
  {{Tarucha}}}]{Matsuo_2023_ArXiv_molecule}%
  \BibitemOpen
  \bibfield  {author} {\bibinfo {author} {\bibfnamefont {S.}~\bibnamefont
  {{Matsuo}}}, \bibinfo {author} {\bibfnamefont {T.}~\bibnamefont {{Imoto}}},
  \bibinfo {author} {\bibfnamefont {T.}~\bibnamefont {{Yokoyama}}}, \bibinfo
  {author} {\bibfnamefont {Y.}~\bibnamefont {{Sato}}}, \bibinfo {author}
  {\bibfnamefont {T.}~\bibnamefont {{Lindemann}}}, \bibinfo {author}
  {\bibfnamefont {S.}~\bibnamefont {{Gronin}}}, \bibinfo {author}
  {\bibfnamefont {G.~C.}\ \bibnamefont {{Gardner}}}, \bibinfo {author}
  {\bibfnamefont {S.}~\bibnamefont {{Nakosai}}}, \bibinfo {author}
  {\bibfnamefont {Y.}~\bibnamefont {{Tanaka}}}, \bibinfo {author}
  {\bibfnamefont {M.~J.}\ \bibnamefont {{Manfra}}},\ and\ \bibinfo {author}
  {\bibfnamefont {S.}~\bibnamefont {{Tarucha}}},\ }\bibfield  {title} {\bibinfo
  {title} {{Phase-dependent Andreev molecules and superconducting gap closing
  in coherently coupled Josephson junctions}},\ }\Eprint
  {https://arxiv.org/abs/http://arxiv.org/abs/2303.10540}
  {http://arxiv.org/abs/2303.10540}  (\bibinfo {year} {2023})\BibitemShut
  {NoStop}%
\bibitem [{\citenamefont {Coraiola}\ \emph
  {et~al.}(2023{\natexlab{b}})\citenamefont {Coraiola}, \citenamefont {Haxell},
  \citenamefont {Sabonis}, \citenamefont {Hinderling}, \citenamefont {ten
  Kate}, \citenamefont {Cheah}, \citenamefont {Krizek}, \citenamefont {Schott},
  \citenamefont {Wegscheider},\ and\ \citenamefont
  {Nichele}}]{Coraiola_2023_ArXiv_spin}%
  \BibitemOpen
  \bibfield  {author} {\bibinfo {author} {\bibfnamefont {M.}~\bibnamefont
  {Coraiola}}, \bibinfo {author} {\bibfnamefont {D.~Z.}\ \bibnamefont
  {Haxell}}, \bibinfo {author} {\bibfnamefont {D.}~\bibnamefont {Sabonis}},
  \bibinfo {author} {\bibfnamefont {M.}~\bibnamefont {Hinderling}}, \bibinfo
  {author} {\bibfnamefont {S.~C.}\ \bibnamefont {ten Kate}}, \bibinfo {author}
  {\bibfnamefont {E.}~\bibnamefont {Cheah}}, \bibinfo {author} {\bibfnamefont
  {F.}~\bibnamefont {Krizek}}, \bibinfo {author} {\bibfnamefont
  {R.}~\bibnamefont {Schott}}, \bibinfo {author} {\bibfnamefont
  {W.}~\bibnamefont {Wegscheider}},\ and\ \bibinfo {author} {\bibfnamefont
  {F.}~\bibnamefont {Nichele}},\ }\bibfield  {title} {\bibinfo {title}
  {{Spin-degeneracy breaking and parity transitions in three-terminal Josephson
  junctions}},\ }\Eprint
  {https://arxiv.org/abs/http://arxiv.org/abs/2307.06715}
  {http://arxiv.org/abs/2307.06715}  (\bibinfo {year}
  {2023}{\natexlab{b}})\BibitemShut {NoStop}%
\bibitem [{\citenamefont {Buzdin}(2008)}]{Buzdin_2008}%
  \BibitemOpen
  \bibfield  {author} {\bibinfo {author} {\bibfnamefont {A.}~\bibnamefont
  {Buzdin}},\ }\bibfield  {title} {\bibinfo {title} {Direct coupling between
  magnetism and superconducting current in the josephson
  ${\ensuremath{\varphi}}_{0}$ junction},\ }\href
  {https://doi.org/10.1103/PhysRevLett.101.107005} {\bibfield  {journal}
  {\bibinfo  {journal} {Phys. Rev. Lett.}\ }\textbf {\bibinfo {volume} {101}},\
  \bibinfo {pages} {107005} (\bibinfo {year} {2008})}\BibitemShut {NoStop}%
\bibitem [{\citenamefont {Haxell}\ \emph
  {et~al.}(2023{\natexlab{a}})\citenamefont {Haxell}, \citenamefont {Coraiola},
  \citenamefont {Hinderling}, \citenamefont {ten Kate}, \citenamefont
  {Sabonis}, \citenamefont {Svetogorov}, \citenamefont {Belzig}, \citenamefont
  {Cheah}, \citenamefont {Krizek}, \citenamefont {Schott}, \citenamefont
  {Wegscheider},\ and\ \citenamefont {Nichele}}]{Haxell_2023}%
  \BibitemOpen
  \bibfield  {author} {\bibinfo {author} {\bibfnamefont {D.~Z.}\ \bibnamefont
  {Haxell}}, \bibinfo {author} {\bibfnamefont {M.}~\bibnamefont {Coraiola}},
  \bibinfo {author} {\bibfnamefont {M.}~\bibnamefont {Hinderling}}, \bibinfo
  {author} {\bibfnamefont {S.~C.}\ \bibnamefont {ten Kate}}, \bibinfo {author}
  {\bibfnamefont {D.}~\bibnamefont {Sabonis}}, \bibinfo {author} {\bibfnamefont
  {A.~E.}\ \bibnamefont {Svetogorov}}, \bibinfo {author} {\bibfnamefont
  {W.}~\bibnamefont {Belzig}}, \bibinfo {author} {\bibfnamefont
  {E.}~\bibnamefont {Cheah}}, \bibinfo {author} {\bibfnamefont
  {F.}~\bibnamefont {Krizek}}, \bibinfo {author} {\bibfnamefont
  {R.}~\bibnamefont {Schott}}, \bibinfo {author} {\bibfnamefont
  {W.}~\bibnamefont {Wegscheider}},\ and\ \bibinfo {author} {\bibfnamefont
  {F.}~\bibnamefont {Nichele}},\ }\bibfield  {title} {\bibinfo {title}
  {{Demonstration of the Nonlocal Josephson Effect in Andreev Molecules}},\
  }\href {https://doi.org/10.1021/acs.nanolett.3c02066} {\bibfield  {journal}
  {\bibinfo  {journal} {Nano Letters}\ }\textbf {\bibinfo {volume} {23}},\
  \bibinfo {pages} {7532} (\bibinfo {year} {2023}{\natexlab{a}})}\BibitemShut
  {NoStop}%
\bibitem [{\citenamefont {Matsuo}\ \emph {et~al.}(2023)\citenamefont {Matsuo},
  \citenamefont {Imoto}, \citenamefont {Yokoyama}, \citenamefont {Sato},
  \citenamefont {Lindemann}, \citenamefont {Gronin}, \citenamefont {Gardner},
  \citenamefont {Manfra},\ and\ \citenamefont
  {Tarucha}}]{Matsuo_2023_ArXiv_nonlocal}%
  \BibitemOpen
  \bibfield  {author} {\bibinfo {author} {\bibfnamefont {S.}~\bibnamefont
  {Matsuo}}, \bibinfo {author} {\bibfnamefont {T.}~\bibnamefont {Imoto}},
  \bibinfo {author} {\bibfnamefont {T.}~\bibnamefont {Yokoyama}}, \bibinfo
  {author} {\bibfnamefont {Y.}~\bibnamefont {Sato}}, \bibinfo {author}
  {\bibfnamefont {T.}~\bibnamefont {Lindemann}}, \bibinfo {author}
  {\bibfnamefont {S.}~\bibnamefont {Gronin}}, \bibinfo {author} {\bibfnamefont
  {G.~C.}\ \bibnamefont {Gardner}}, \bibinfo {author} {\bibfnamefont {M.~J.}\
  \bibnamefont {Manfra}},\ and\ \bibinfo {author} {\bibfnamefont
  {S.}~\bibnamefont {Tarucha}},\ }\bibfield  {title} {\bibinfo {title}
  {{Engineering of anomalous Josephson effect in coherently coupled Josephson
  junctions}},\ }\Eprint
  {https://arxiv.org/abs/http://arxiv.org/abs/2305.06596}
  {http://arxiv.org/abs/2305.06596}  (\bibinfo {year} {2023})\BibitemShut
  {NoStop}%
\bibitem [{\citenamefont {Moehle}\ \emph {et~al.}(2021)\citenamefont {Moehle},
  \citenamefont {Ke}, \citenamefont {Wang}, \citenamefont {Thomas},
  \citenamefont {Xiao}, \citenamefont {Karwal}, \citenamefont {Lodari},
  \citenamefont {van~de Kerkhof}, \citenamefont {Termaat}, \citenamefont
  {Gardner}, \citenamefont {Scappucci}, \citenamefont {Manfra},\ and\
  \citenamefont {Goswami}}]{Moehle_2021}%
  \BibitemOpen
  \bibfield  {author} {\bibinfo {author} {\bibfnamefont {C.~M.}\ \bibnamefont
  {Moehle}}, \bibinfo {author} {\bibfnamefont {C.~T.}\ \bibnamefont {Ke}},
  \bibinfo {author} {\bibfnamefont {Q.}~\bibnamefont {Wang}}, \bibinfo {author}
  {\bibfnamefont {C.}~\bibnamefont {Thomas}}, \bibinfo {author} {\bibfnamefont
  {D.}~\bibnamefont {Xiao}}, \bibinfo {author} {\bibfnamefont {S.}~\bibnamefont
  {Karwal}}, \bibinfo {author} {\bibfnamefont {M.}~\bibnamefont {Lodari}},
  \bibinfo {author} {\bibfnamefont {V.}~\bibnamefont {van~de Kerkhof}},
  \bibinfo {author} {\bibfnamefont {R.}~\bibnamefont {Termaat}}, \bibinfo
  {author} {\bibfnamefont {G.~C.}\ \bibnamefont {Gardner}}, \bibinfo {author}
  {\bibfnamefont {G.}~\bibnamefont {Scappucci}}, \bibinfo {author}
  {\bibfnamefont {M.~J.}\ \bibnamefont {Manfra}},\ and\ \bibinfo {author}
  {\bibfnamefont {S.}~\bibnamefont {Goswami}},\ }\bibfield  {title} {\bibinfo
  {title} {{{InSbAs} Two-Dimensional Electron Gases as a Platform for
  Topological Superconductivity}},\ }\href
  {https://doi.org/10.1021/acs.nanolett.1c03520} {\bibfield  {journal}
  {\bibinfo  {journal} {Nano Letters}\ }\textbf {\bibinfo {volume} {21}},\
  \bibinfo {pages} {9990} (\bibinfo {year} {2021})}\BibitemShut {NoStop}%
\bibitem [{\citenamefont {Miyazaki}\ \emph {et~al.}(2006)\citenamefont
  {Miyazaki}, \citenamefont {Kanda},\ and\ \citenamefont
  {Ootuka}}]{Miyazaki_2006}%
  \BibitemOpen
  \bibfield  {author} {\bibinfo {author} {\bibfnamefont {H.}~\bibnamefont
  {Miyazaki}}, \bibinfo {author} {\bibfnamefont {A.}~\bibnamefont {Kanda}},\
  and\ \bibinfo {author} {\bibfnamefont {Y.}~\bibnamefont {Ootuka}},\
  }\bibfield  {title} {\bibinfo {title} {{Current-phase relation of a
  superconducting quantum point contact}},\ }\href
  {https://doi.org/10.1016/j.physc.2005.12.051} {\bibfield  {journal} {\bibinfo
   {journal} {Physica C: Superconductivity and its Applications}\ }\textbf
  {\bibinfo {volume} {437-438}},\ \bibinfo {pages} {217} (\bibinfo {year}
  {2006})}\BibitemShut {NoStop}%
\bibitem [{\citenamefont {Rocca}\ \emph {et~al.}(2007)\citenamefont {Rocca},
  \citenamefont {Chauvin}, \citenamefont {Huard}, \citenamefont {Pothier},
  \citenamefont {Esteve},\ and\ \citenamefont {Urbina}}]{Della_Rocca_2007}%
  \BibitemOpen
  \bibfield  {author} {\bibinfo {author} {\bibfnamefont {M.~L.~D.}\
  \bibnamefont {Rocca}}, \bibinfo {author} {\bibfnamefont {M.}~\bibnamefont
  {Chauvin}}, \bibinfo {author} {\bibfnamefont {B.}~\bibnamefont {Huard}},
  \bibinfo {author} {\bibfnamefont {H.}~\bibnamefont {Pothier}}, \bibinfo
  {author} {\bibfnamefont {D.}~\bibnamefont {Esteve}},\ and\ \bibinfo {author}
  {\bibfnamefont {C.}~\bibnamefont {Urbina}},\ }\bibfield  {title} {\bibinfo
  {title} {{Measurement of the Current-Phase Relation of Superconducting Atomic
  Contacts}},\ }\bibfield  {journal} {\bibinfo  {journal} {Physical Review
  Letters}\ }\textbf {\bibinfo {volume} {99}},\ \href
  {https://doi.org/10.1103/physrevlett.99.127005}
  {10.1103/physrevlett.99.127005} (\bibinfo {year} {2007})\BibitemShut
  {NoStop}%
\bibitem [{Note1()}]{Note1}%
  \BibitemOpen
  \bibinfo {note} {The probes are floating for all measurements except those of
  Figs.~\protect \hyperref [fig:4termjj_characterization]{\ref
  *{fig:4termjj_characterization}(d)}, \protect \hyperref
  [fig:4termjj_anomalous]{\ref *{fig:4termjj_anomalous}(c)}, and \ref
  {fig:4termjj_gate_dependence}, where one is grounded. As switching current
  measurements, these measurements are unaffected by the grounded probes, since
  current favors traveling through the superconducting circuit until the
  current bias is large enough that the circuit switches into a resistive
  state.}\BibitemShut {Stop}%
\bibitem [{\citenamefont {Prosko}\ \emph {et~al.}(2023)\citenamefont {Prosko},
  \citenamefont {Kulesh}, \citenamefont {Chan}, \citenamefont {Han},
  \citenamefont {Xiao}, \citenamefont {Thomas}, \citenamefont {Manfra},
  \citenamefont {Goswami},\ and\ \citenamefont
  {Malinowski}}]{Prosko_2023_ArXiv}%
  \BibitemOpen
  \bibfield  {author} {\bibinfo {author} {\bibfnamefont {C.~G.}\ \bibnamefont
  {Prosko}}, \bibinfo {author} {\bibfnamefont {I.}~\bibnamefont {Kulesh}},
  \bibinfo {author} {\bibfnamefont {M.}~\bibnamefont {Chan}}, \bibinfo {author}
  {\bibfnamefont {L.}~\bibnamefont {Han}}, \bibinfo {author} {\bibfnamefont
  {D.}~\bibnamefont {Xiao}}, \bibinfo {author} {\bibfnamefont {C.}~\bibnamefont
  {Thomas}}, \bibinfo {author} {\bibfnamefont {M.~J.}\ \bibnamefont {Manfra}},
  \bibinfo {author} {\bibfnamefont {S.}~\bibnamefont {Goswami}},\ and\ \bibinfo
  {author} {\bibfnamefont {F.~K.}\ \bibnamefont {Malinowski}},\ }\bibfield
  {title} {\bibinfo {title} {{Flux-Tunable Hybridization in a Double Quantum
  Dot Interferometer}},\ }\Eprint {https://arxiv.org/abs/2303.04144}
  {2303.04144}  (\bibinfo {year} {2023})\BibitemShut {NoStop}%
\bibitem [{\citenamefont {Nanda}\ \emph {et~al.}(2017)\citenamefont {Nanda},
  \citenamefont {Aguilera-Servin}, \citenamefont {Rakyta}, \citenamefont
  {Korm{\'{a}}nyos}, \citenamefont {Kleiner}, \citenamefont {Koelle},
  \citenamefont {Watanabe}, \citenamefont {Taniguchi}, \citenamefont
  {Vandersypen},\ and\ \citenamefont {Goswami}}]{Nanda_2017}%
  \BibitemOpen
  \bibfield  {author} {\bibinfo {author} {\bibfnamefont {G.}~\bibnamefont
  {Nanda}}, \bibinfo {author} {\bibfnamefont {J.~L.}\ \bibnamefont
  {Aguilera-Servin}}, \bibinfo {author} {\bibfnamefont {P.}~\bibnamefont
  {Rakyta}}, \bibinfo {author} {\bibfnamefont {A.}~\bibnamefont
  {Korm{\'{a}}nyos}}, \bibinfo {author} {\bibfnamefont {R.}~\bibnamefont
  {Kleiner}}, \bibinfo {author} {\bibfnamefont {D.}~\bibnamefont {Koelle}},
  \bibinfo {author} {\bibfnamefont {K.}~\bibnamefont {Watanabe}}, \bibinfo
  {author} {\bibfnamefont {T.}~\bibnamefont {Taniguchi}}, \bibinfo {author}
  {\bibfnamefont {L.~M.~K.}\ \bibnamefont {Vandersypen}},\ and\ \bibinfo
  {author} {\bibfnamefont {S.}~\bibnamefont {Goswami}},\ }\bibfield  {title}
  {\bibinfo {title} {{Current-Phase Relation of Ballistic Graphene Josephson
  Junctions}},\ }\href {https://doi.org/10.1021/acs.nanolett.7b00097}
  {\bibfield  {journal} {\bibinfo  {journal} {Nano Letters}\ }\textbf {\bibinfo
  {volume} {17}},\ \bibinfo {pages} {3396} (\bibinfo {year}
  {2017})}\BibitemShut {NoStop}%
\bibitem [{\citenamefont {Tinkham}(1996)}]{Tinkham}%
  \BibitemOpen
  \bibfield  {author} {\bibinfo {author} {\bibfnamefont {M.}~\bibnamefont
  {Tinkham}},\ }\href@noop {} {\emph {\bibinfo {title} {{Introduction to
  superconductivity}}}},\ \bibinfo {edition} {2nd}\ ed.\ (\bibinfo  {publisher}
  {Dover publications, inc.},\ \bibinfo {year} {1996})\BibitemShut {NoStop}%
\bibitem [{\citenamefont {Haxell}\ \emph
  {et~al.}(2023{\natexlab{b}})\citenamefont {Haxell}, \citenamefont {Cheah},
  \citenamefont {K{\v{r}}{\'{\i}}{\v{z}}ek}, \citenamefont {Schott},
  \citenamefont {Ritter}, \citenamefont {Hinderling}, \citenamefont {Belzig},
  \citenamefont {Bruder}, \citenamefont {Wegscheider}, \citenamefont {Riel},\
  and\ \citenamefont {Nichele}}]{Haxell_2023_switch}%
  \BibitemOpen
  \bibfield  {author} {\bibinfo {author} {\bibfnamefont {D.}~\bibnamefont
  {Haxell}}, \bibinfo {author} {\bibfnamefont {E.}~\bibnamefont {Cheah}},
  \bibinfo {author} {\bibfnamefont {F.}~\bibnamefont
  {K{\v{r}}{\'{\i}}{\v{z}}ek}}, \bibinfo {author} {\bibfnamefont
  {R.}~\bibnamefont {Schott}}, \bibinfo {author} {\bibfnamefont
  {M.}~\bibnamefont {Ritter}}, \bibinfo {author} {\bibfnamefont
  {M.}~\bibnamefont {Hinderling}}, \bibinfo {author} {\bibfnamefont
  {W.}~\bibnamefont {Belzig}}, \bibinfo {author} {\bibfnamefont
  {C.}~\bibnamefont {Bruder}}, \bibinfo {author} {\bibfnamefont
  {W.}~\bibnamefont {Wegscheider}}, \bibinfo {author} {\bibfnamefont
  {H.}~\bibnamefont {Riel}},\ and\ \bibinfo {author} {\bibfnamefont
  {F.}~\bibnamefont {Nichele}},\ }\bibfield  {title} {\bibinfo {title}
  {{Measurements of Phase Dynamics in Planar Josephson Junctions and
  {SQUIDs}}},\ }\bibfield  {journal} {\bibinfo  {journal} {Physical Review
  Letters}\ }\textbf {\bibinfo {volume} {130}},\ \href
  {https://doi.org/10.1103/physrevlett.130.087002}
  {10.1103/physrevlett.130.087002} (\bibinfo {year}
  {2023}{\natexlab{b}})\BibitemShut {NoStop}%
\bibitem [{\citenamefont {Matsuo}\ \emph {et~al.}(2022)\citenamefont {Matsuo},
  \citenamefont {Lee}, \citenamefont {Chang}, \citenamefont {Sato},
  \citenamefont {Ueda}, \citenamefont {Palmstr{\o}m},\ and\ \citenamefont
  {Tarucha}}]{Matsuo_2022}%
  \BibitemOpen
  \bibfield  {author} {\bibinfo {author} {\bibfnamefont {S.}~\bibnamefont
  {Matsuo}}, \bibinfo {author} {\bibfnamefont {J.~S.}\ \bibnamefont {Lee}},
  \bibinfo {author} {\bibfnamefont {C.-Y.}\ \bibnamefont {Chang}}, \bibinfo
  {author} {\bibfnamefont {Y.}~\bibnamefont {Sato}}, \bibinfo {author}
  {\bibfnamefont {K.}~\bibnamefont {Ueda}}, \bibinfo {author} {\bibfnamefont
  {C.~J.}\ \bibnamefont {Palmstr{\o}m}},\ and\ \bibinfo {author} {\bibfnamefont
  {S.}~\bibnamefont {Tarucha}},\ }\bibfield  {title} {\bibinfo {title}
  {{Observation of nonlocal Josephson effect on double {InAs} nanowires}},\
  }\bibfield  {journal} {\bibinfo  {journal} {Communications Physics}\ }\textbf
  {\bibinfo {volume} {5}},\ \href {https://doi.org/10.1038/s42005-022-00994-0}
  {10.1038/s42005-022-00994-0} (\bibinfo {year} {2022})\BibitemShut {NoStop}%
\bibitem [{\citenamefont {Strambini}\ \emph {et~al.}(2020)\citenamefont
  {Strambini}, \citenamefont {Iorio}, \citenamefont {Durante}, \citenamefont
  {Citro}, \citenamefont {Sanz-Fern{\'{a}}ndez}, \citenamefont {Guarcello},
  \citenamefont {Tokatly}, \citenamefont {Braggio}, \citenamefont {Rocci},
  \citenamefont {Ligato}, \citenamefont {Zannier}, \citenamefont {Sorba},
  \citenamefont {Bergeret},\ and\ \citenamefont {Giazotto}}]{Strambini_2020}%
  \BibitemOpen
  \bibfield  {author} {\bibinfo {author} {\bibfnamefont {E.}~\bibnamefont
  {Strambini}}, \bibinfo {author} {\bibfnamefont {A.}~\bibnamefont {Iorio}},
  \bibinfo {author} {\bibfnamefont {O.}~\bibnamefont {Durante}}, \bibinfo
  {author} {\bibfnamefont {R.}~\bibnamefont {Citro}}, \bibinfo {author}
  {\bibfnamefont {C.}~\bibnamefont {Sanz-Fern{\'{a}}ndez}}, \bibinfo {author}
  {\bibfnamefont {C.}~\bibnamefont {Guarcello}}, \bibinfo {author}
  {\bibfnamefont {I.~V.}\ \bibnamefont {Tokatly}}, \bibinfo {author}
  {\bibfnamefont {A.}~\bibnamefont {Braggio}}, \bibinfo {author} {\bibfnamefont
  {M.}~\bibnamefont {Rocci}}, \bibinfo {author} {\bibfnamefont
  {N.}~\bibnamefont {Ligato}}, \bibinfo {author} {\bibfnamefont
  {V.}~\bibnamefont {Zannier}}, \bibinfo {author} {\bibfnamefont
  {L.}~\bibnamefont {Sorba}}, \bibinfo {author} {\bibfnamefont {F.~S.}\
  \bibnamefont {Bergeret}},\ and\ \bibinfo {author} {\bibfnamefont
  {F.}~\bibnamefont {Giazotto}},\ }\bibfield  {title} {\bibinfo {title} {{A
  Josephson phase battery}},\ }\href
  {https://doi.org/10.1038/s41565-020-0712-7} {\bibfield  {journal} {\bibinfo
  {journal} {Nature Nanotechnology}\ }\textbf {\bibinfo {volume} {15}},\
  \bibinfo {pages} {656} (\bibinfo {year} {2020})}\BibitemShut {NoStop}%
\bibitem [{\citenamefont {Pillet}\ \emph {et~al.}(2019)\citenamefont {Pillet},
  \citenamefont {Benzoni}, \citenamefont {Griesmar}, \citenamefont {Smirr},\
  and\ \citenamefont {Girit}}]{Pillet_2019}%
  \BibitemOpen
  \bibfield  {author} {\bibinfo {author} {\bibfnamefont {J.-D.}\ \bibnamefont
  {Pillet}}, \bibinfo {author} {\bibfnamefont {V.}~\bibnamefont {Benzoni}},
  \bibinfo {author} {\bibfnamefont {J.}~\bibnamefont {Griesmar}}, \bibinfo
  {author} {\bibfnamefont {J.-L.}\ \bibnamefont {Smirr}},\ and\ \bibinfo
  {author} {\bibfnamefont {{\c{C}}.~{\"{O}}.}\ \bibnamefont {Girit}},\
  }\bibfield  {title} {\bibinfo {title} {{Nonlocal Josephson Effect in Andreev
  Molecules}},\ }\href {https://doi.org/10.1021/acs.nanolett.9b02686}
  {\bibfield  {journal} {\bibinfo  {journal} {Nano Letters}\ }\textbf {\bibinfo
  {volume} {19}},\ \bibinfo {pages} {7138} (\bibinfo {year}
  {2019})}\BibitemShut {NoStop}%
\bibitem [{\citenamefont {Kocsis}\ \emph {et~al.}(2023)\citenamefont {Kocsis},
  \citenamefont {Scherübl}, \citenamefont {Fülöp}, \citenamefont {Makk},\
  and\ \citenamefont {Csonka}}]{Kocsis_2023_ArXiv}%
  \BibitemOpen
  \bibfield  {author} {\bibinfo {author} {\bibfnamefont {M.}~\bibnamefont
  {Kocsis}}, \bibinfo {author} {\bibfnamefont {Z.}~\bibnamefont {Scherübl}},
  \bibinfo {author} {\bibfnamefont {G.}~\bibnamefont {Fülöp}}, \bibinfo
  {author} {\bibfnamefont {P.}~\bibnamefont {Makk}},\ and\ \bibinfo {author}
  {\bibfnamefont {S.}~\bibnamefont {Csonka}},\ }\bibfield  {title} {\bibinfo
  {title} {{Strong nonlocal tuning of the current-phase relation of a quantum
  dot based Andreev molecule}},\ }\Eprint
  {https://arxiv.org/abs/http://arxiv.org/abs/2303.14842}
  {http://arxiv.org/abs/2303.14842}  (\bibinfo {year} {2023})\BibitemShut
  {NoStop}%
\bibitem [{\citenamefont {Rasmussen}\ \emph {et~al.}(2021)\citenamefont
  {Rasmussen}, \citenamefont {Christensen}, \citenamefont {Pedersen},
  \citenamefont {Kristensen}, \citenamefont {B{\ae}kkegaard}, \citenamefont
  {Loft},\ and\ \citenamefont {Zinner}}]{Rasmussen_2021}%
  \BibitemOpen
  \bibfield  {author} {\bibinfo {author} {\bibfnamefont {S.}~\bibnamefont
  {Rasmussen}}, \bibinfo {author} {\bibfnamefont {K.}~\bibnamefont
  {Christensen}}, \bibinfo {author} {\bibfnamefont {S.}~\bibnamefont
  {Pedersen}}, \bibinfo {author} {\bibfnamefont {L.}~\bibnamefont
  {Kristensen}}, \bibinfo {author} {\bibfnamefont {T.}~\bibnamefont
  {B{\ae}kkegaard}}, \bibinfo {author} {\bibfnamefont {N.}~\bibnamefont
  {Loft}},\ and\ \bibinfo {author} {\bibfnamefont {N.}~\bibnamefont {Zinner}},\
  }\bibfield  {title} {\bibinfo {title} {{Superconducting Circuit
  Companion{\textemdash}an Introduction with Worked Examples}},\ }\bibfield
  {journal} {\bibinfo  {journal} {{PRX} Quantum}\ }\textbf {\bibinfo {volume}
  {2}},\ \href {https://doi.org/10.1103/prxquantum.2.040204}
  {10.1103/prxquantum.2.040204} (\bibinfo {year} {2021})\BibitemShut {NoStop}%
\bibitem [{\citenamefont {Kornich}\ \emph {et~al.}(2020)\citenamefont
  {Kornich}, \citenamefont {Barakov},\ and\ \citenamefont
  {Nazarov}}]{Kornich_2020}%
  \BibitemOpen
  \bibfield  {author} {\bibinfo {author} {\bibfnamefont {V.}~\bibnamefont
  {Kornich}}, \bibinfo {author} {\bibfnamefont {H.~S.}\ \bibnamefont
  {Barakov}},\ and\ \bibinfo {author} {\bibfnamefont {Y.~V.}\ \bibnamefont
  {Nazarov}},\ }\bibfield  {title} {\bibinfo {title} {Overlapping andreev
  states in semiconducting nanowires: Competition of one-dimensional and
  three-dimensional propagation},\ }\bibfield  {journal} {\bibinfo  {journal}
  {Physical Review B}\ }\textbf {\bibinfo {volume} {101}},\ \href
  {https://doi.org/10.1103/physrevb.101.195430} {10.1103/physrevb.101.195430}
  (\bibinfo {year} {2020})\BibitemShut {NoStop}%
\bibitem [{\citenamefont {Kürtössy}\ \emph {et~al.}(2021)\citenamefont
  {Kürtössy}, \citenamefont {Scherübl}, \citenamefont {Fülöp},
  \citenamefont {Luk{\'{a}}cs}, \citenamefont {Kanne}, \citenamefont
  {Nyg{\aa}rd}, \citenamefont {Makk},\ and\ \citenamefont
  {Csonka}}]{Kurtossy_2021}%
  \BibitemOpen
  \bibfield  {author} {\bibinfo {author} {\bibfnamefont {O.}~\bibnamefont
  {Kürtössy}}, \bibinfo {author} {\bibfnamefont {Z.}~\bibnamefont
  {Scherübl}}, \bibinfo {author} {\bibfnamefont {G.}~\bibnamefont {Fülöp}},
  \bibinfo {author} {\bibfnamefont {I.~E.}\ \bibnamefont {Luk{\'{a}}cs}},
  \bibinfo {author} {\bibfnamefont {T.}~\bibnamefont {Kanne}}, \bibinfo
  {author} {\bibfnamefont {J.}~\bibnamefont {Nyg{\aa}rd}}, \bibinfo {author}
  {\bibfnamefont {P.}~\bibnamefont {Makk}},\ and\ \bibinfo {author}
  {\bibfnamefont {S.}~\bibnamefont {Csonka}},\ }\bibfield  {title} {\bibinfo
  {title} {{Andreev Molecule in Parallel {InAs} Nanowires}},\ }\href
  {https://doi.org/10.1021/acs.nanolett.1c01956} {\bibfield  {journal}
  {\bibinfo  {journal} {Nano Letters}\ }\textbf {\bibinfo {volume} {21}},\
  \bibinfo {pages} {7929} (\bibinfo {year} {2021})}\BibitemShut {NoStop}%
\bibitem [{\citenamefont {Pita-Vidal}\ \emph {et~al.}(2023)\citenamefont
  {Pita-Vidal}, \citenamefont {Wesdorp}, \citenamefont {Splitthoff},
  \citenamefont {Bargerbos}, \citenamefont {Liu}, \citenamefont {Kouwenhoven},\
  and\ \citenamefont {Andersen}}]{PitaVidal_2023_ArXiv}%
  \BibitemOpen
  \bibfield  {author} {\bibinfo {author} {\bibfnamefont {M.}~\bibnamefont
  {Pita-Vidal}}, \bibinfo {author} {\bibfnamefont {J.~J.}\ \bibnamefont
  {Wesdorp}}, \bibinfo {author} {\bibfnamefont {L.~J.}\ \bibnamefont
  {Splitthoff}}, \bibinfo {author} {\bibfnamefont {A.}~\bibnamefont
  {Bargerbos}}, \bibinfo {author} {\bibfnamefont {Y.}~\bibnamefont {Liu}},
  \bibinfo {author} {\bibfnamefont {L.~P.}\ \bibnamefont {Kouwenhoven}},\ and\
  \bibinfo {author} {\bibfnamefont {C.~K.}\ \bibnamefont {Andersen}},\
  }\bibfield  {title} {\bibinfo {title} {{Strong tunable coupling between two
  distant superconducting spin qubits}},\ }\Eprint
  {https://arxiv.org/abs/http://arxiv.org/abs/2307.15654}
  {http://arxiv.org/abs/2307.15654}  (\bibinfo {year} {2023})\BibitemShut
  {NoStop}%
\bibitem [{\citenamefont {Cheung}\ \emph {et~al.}(2023)\citenamefont {Cheung},
  \citenamefont {Haller}, \citenamefont {Kononov}, \citenamefont {Ciaccia},
  \citenamefont {Ungerer}, \citenamefont {Kanne}, \citenamefont {Nygård},
  \citenamefont {Winkel}, \citenamefont {Reisinger}, \citenamefont {Pop},
  \citenamefont {Baumgartner},\ and\ \citenamefont
  {Schönenberger}}]{Cheung_2023_ArXiv}%
  \BibitemOpen
  \bibfield  {author} {\bibinfo {author} {\bibfnamefont {L.~Y.}\ \bibnamefont
  {Cheung}}, \bibinfo {author} {\bibfnamefont {R.}~\bibnamefont {Haller}},
  \bibinfo {author} {\bibfnamefont {A.}~\bibnamefont {Kononov}}, \bibinfo
  {author} {\bibfnamefont {C.}~\bibnamefont {Ciaccia}}, \bibinfo {author}
  {\bibfnamefont {J.~H.}\ \bibnamefont {Ungerer}}, \bibinfo {author}
  {\bibfnamefont {T.}~\bibnamefont {Kanne}}, \bibinfo {author} {\bibfnamefont
  {J.}~\bibnamefont {Nygård}}, \bibinfo {author} {\bibfnamefont
  {P.}~\bibnamefont {Winkel}}, \bibinfo {author} {\bibfnamefont
  {T.}~\bibnamefont {Reisinger}}, \bibinfo {author} {\bibfnamefont {I.~M.}\
  \bibnamefont {Pop}}, \bibinfo {author} {\bibfnamefont {A.}~\bibnamefont
  {Baumgartner}},\ and\ \bibinfo {author} {\bibfnamefont {C.}~\bibnamefont
  {Schönenberger}},\ }\bibfield  {title} {\bibinfo {title} {{Photon-mediated
  long range coupling of two Andreev level qubits}},\ }\Eprint
  {https://arxiv.org/abs/2310.15995} {2310.15995}  (\bibinfo {year}
  {2023})\BibitemShut {NoStop}%
\bibitem [{\citenamefont {Geier}\ \emph {et~al.}(2023)\citenamefont {Geier},
  \citenamefont {Souto}, \citenamefont {Schulenborg}, \citenamefont {Asaad},
  \citenamefont {Leijnse},\ and\ \citenamefont {Flensberg}}]{Geier_2023_ArXiv}%
  \BibitemOpen
  \bibfield  {author} {\bibinfo {author} {\bibfnamefont {M.}~\bibnamefont
  {Geier}}, \bibinfo {author} {\bibfnamefont {R.~S.}\ \bibnamefont {Souto}},
  \bibinfo {author} {\bibfnamefont {J.}~\bibnamefont {Schulenborg}}, \bibinfo
  {author} {\bibfnamefont {S.}~\bibnamefont {Asaad}}, \bibinfo {author}
  {\bibfnamefont {M.}~\bibnamefont {Leijnse}},\ and\ \bibinfo {author}
  {\bibfnamefont {K.}~\bibnamefont {Flensberg}},\ }\bibfield  {title} {\bibinfo
  {title} {A fermion-parity qubit in a proximitized double quantum dot},\
  }\Eprint {https://arxiv.org/abs/2307.05678} {2307.05678}  (\bibinfo {year}
  {2023})\BibitemShut {NoStop}%
\bibitem [{\citenamefont {Fulga}\ \emph {et~al.}(2013)\citenamefont {Fulga},
  \citenamefont {Haim}, \citenamefont {Akhmerov},\ and\ \citenamefont
  {Oreg}}]{Fulga_2013}%
  \BibitemOpen
  \bibfield  {author} {\bibinfo {author} {\bibfnamefont {I.~C.}\ \bibnamefont
  {Fulga}}, \bibinfo {author} {\bibfnamefont {A.}~\bibnamefont {Haim}},
  \bibinfo {author} {\bibfnamefont {A.~R.}\ \bibnamefont {Akhmerov}},\ and\
  \bibinfo {author} {\bibfnamefont {Y.}~\bibnamefont {Oreg}},\ }\bibfield
  {title} {\bibinfo {title} {{Adaptive tuning of Majorana fermions in a quantum
  dot chain}},\ }\href {https://doi.org/10.1088/1367-2630/15/4/045020}
  {\bibfield  {journal} {\bibinfo  {journal} {New Journal of Physics}\ }\textbf
  {\bibinfo {volume} {15}},\ \bibinfo {pages} {045020} (\bibinfo {year}
  {2013})}\BibitemShut {NoStop}%
\bibitem [{\citenamefont {Zazunov}\ \emph {et~al.}(2003)\citenamefont
  {Zazunov}, \citenamefont {Shumeiko}, \citenamefont {Bratus'}, \citenamefont
  {Lantz},\ and\ \citenamefont {Wendin}}]{Zazunov_2003}%
  \BibitemOpen
  \bibfield  {author} {\bibinfo {author} {\bibfnamefont {A.}~\bibnamefont
  {Zazunov}}, \bibinfo {author} {\bibfnamefont {V.~S.}\ \bibnamefont
  {Shumeiko}}, \bibinfo {author} {\bibfnamefont {E.~N.}\ \bibnamefont
  {Bratus'}}, \bibinfo {author} {\bibfnamefont {J.}~\bibnamefont {Lantz}},\
  and\ \bibinfo {author} {\bibfnamefont {G.}~\bibnamefont {Wendin}},\
  }\bibfield  {title} {\bibinfo {title} {Andreev level qubit},\ }\bibfield
  {journal} {\bibinfo  {journal} {Physical Review Letters}\ }\textbf {\bibinfo
  {volume} {90}},\ \href {https://doi.org/10.1103/physrevlett.90.087003}
  {10.1103/physrevlett.90.087003} (\bibinfo {year} {2003})\BibitemShut
  {NoStop}%
\bibitem [{\citenamefont {Janvier}\ \emph {et~al.}(2015)\citenamefont
  {Janvier}, \citenamefont {Tosi}, \citenamefont {Bretheau}, \citenamefont
  {Girit}, \citenamefont {Stern}, \citenamefont {Bertet}, \citenamefont
  {Joyez}, \citenamefont {Vion}, \citenamefont {Esteve}, \citenamefont
  {Goffman}, \citenamefont {Pothier},\ and\ \citenamefont
  {Urbina}}]{Janvier_2015}%
  \BibitemOpen
  \bibfield  {author} {\bibinfo {author} {\bibfnamefont {C.}~\bibnamefont
  {Janvier}}, \bibinfo {author} {\bibfnamefont {L.}~\bibnamefont {Tosi}},
  \bibinfo {author} {\bibfnamefont {L.}~\bibnamefont {Bretheau}}, \bibinfo
  {author} {\bibfnamefont {{\c{C}}.~O.}\ \bibnamefont {Girit}}, \bibinfo
  {author} {\bibfnamefont {M.}~\bibnamefont {Stern}}, \bibinfo {author}
  {\bibfnamefont {P.}~\bibnamefont {Bertet}}, \bibinfo {author} {\bibfnamefont
  {P.}~\bibnamefont {Joyez}}, \bibinfo {author} {\bibfnamefont
  {D.}~\bibnamefont {Vion}}, \bibinfo {author} {\bibfnamefont {D.}~\bibnamefont
  {Esteve}}, \bibinfo {author} {\bibfnamefont {M.~F.}\ \bibnamefont {Goffman}},
  \bibinfo {author} {\bibfnamefont {H.}~\bibnamefont {Pothier}},\ and\ \bibinfo
  {author} {\bibfnamefont {C.}~\bibnamefont {Urbina}},\ }\bibfield  {title}
  {\bibinfo {title} {Coherent manipulation of andreev states in superconducting
  atomic contacts},\ }\href {https://doi.org/10.1126/science.aab2179}
  {\bibfield  {journal} {\bibinfo  {journal} {Science}\ }\textbf {\bibinfo
  {volume} {349}},\ \bibinfo {pages} {1199} (\bibinfo {year}
  {2015})}\BibitemShut {NoStop}%
\bibitem [{\citenamefont {Xie}\ \emph {et~al.}(2018)\citenamefont {Xie},
  \citenamefont {Vavilov},\ and\ \citenamefont {Levchenko}}]{Xie_2018}%
  \BibitemOpen
  \bibfield  {author} {\bibinfo {author} {\bibfnamefont {H.-Y.}\ \bibnamefont
  {Xie}}, \bibinfo {author} {\bibfnamefont {M.~G.}\ \bibnamefont {Vavilov}},\
  and\ \bibinfo {author} {\bibfnamefont {A.}~\bibnamefont {Levchenko}},\
  }\bibfield  {title} {\bibinfo {title} {Weyl nodes in andreev spectra of
  multiterminal josephson junctions: Chern numbers, conductances, and
  supercurrents},\ }\bibfield  {journal} {\bibinfo  {journal} {Physical Review
  B}\ }\textbf {\bibinfo {volume} {97}},\ \href
  {https://doi.org/10.1103/physrevb.97.035443} {10.1103/physrevb.97.035443}
  (\bibinfo {year} {2018})\BibitemShut {NoStop}%
\bibitem [{\citenamefont {{Christian G. Prosko and Wietze D. Huisman and Ivan
  Kuleshand Di Xiao and Candice Thomas and Michael J. Manfra and Srijit
  Goswami}}(2023)}]{datarepo_4termjj}%
  \BibitemOpen
  \bibfield  {author} {\bibinfo {author} {\bibnamefont {{Christian G. Prosko
  and Wietze D. Huisman and Ivan Kuleshand Di Xiao and Candice Thomas and
  Michael J. Manfra and Srijit Goswami}}},\ }\href
  {https://zenodo.org/doi/10.5281/zenodo.10212998} {\bibinfo {title} {{Data
  repository accompanying ``Flux-tunable Josephson Effect in a Four-Terminal
  Junction''}}},\ \bibinfo {howpublished}
  {https://doi.org/10.5281/zenodo.10212998} (\bibinfo {year}
  {2023})\BibitemShut {NoStop}%
\bibitem [{\citenamefont {Golubov}\ \emph {et~al.}(2004)\citenamefont
  {Golubov}, \citenamefont {Kupriyanov},\ and\ \citenamefont
  {Il'ichev}}]{Golubov_2004}%
  \BibitemOpen
  \bibfield  {author} {\bibinfo {author} {\bibfnamefont {A.~A.}\ \bibnamefont
  {Golubov}}, \bibinfo {author} {\bibfnamefont {M.~Y.}\ \bibnamefont
  {Kupriyanov}},\ and\ \bibinfo {author} {\bibfnamefont {E.}~\bibnamefont
  {Il'ichev}},\ }\bibfield  {title} {\bibinfo {title} {The current-phase
  relation in josephson junctions},\ }\href
  {https://doi.org/10.1103/revmodphys.76.411} {\bibfield  {journal} {\bibinfo
  {journal} {Reviews of Modern Physics}\ }\textbf {\bibinfo {volume} {76}},\
  \bibinfo {pages} {411} (\bibinfo {year} {2004})}\BibitemShut {NoStop}%
\end{thebibliography}%

\end{document}